\documentclass[a4paper,11pt,twoside]{article}
\usepackage{amsmath} 
\usepackage{amssymb} 
\usepackage{amsthm}  
\usepackage{fancyhdr}
\usepackage{enumitem}
\usepackage[utf8]{inputenc}
\usepackage{setspace}
\usepackage{graphicx}
\usepackage{array}

\usepackage{epstopdf}

\usepackage[a4paper,twoside, hmarginratio={1:1}]{geometry}

\newcommand{\1}{\frac{1}{2}}
\renewcommand{\o}[1]{\overline{#1}}

\def\s0{\sigma_0}
\def\seff{\sigma_{\textrm{eff}}}
\def\sol{\sigma_{\textrm{one-loop}}}
\def\saeff{\sigma_{\alpha\textrm{eff}}}
\def\DU{\Delta_{\rm eff}}
\def\DNU{\Delta_{NU}}

\def\DA{\Delta_{\alpha}}

\def\chii{\tilde{\chi}_1^0}

\def\noi{\noindent}

\begin{document}

\def\mhf{M_{1/2}}

\newcommand{\psla}{\mbox{$\not{\! p}$}}
\newcommand{\qsla}{\mbox{$\not{\! q}$}}
\newcommand{\bra}{\langle}
\newcommand{\ket}{\rangle}

\newcommand{\mchi}{{m_{\chi}}}
\newcommand{\neuti}{{\tilde{\chi}}_i^0}
\newcommand{\neutj}{{\tilde{\chi}}_j^0}
\newcommand{\neuto}{{\tilde{\chi}}_1^0}
\newcommand{\mneuto}{m_{{\tilde{\chi}}_1^0}}
\newcommand{\neutt}{{\tilde{\chi}}_2^0}
\newcommand{\mneutt}{m_{{\tilde{\chi}}_2^0}}
\newcommand{\neutth}{{\tilde{\chi}}_3^0}
\newcommand{\mneutth}{m_{{\tilde{\chi}}_3^0}}
\newcommand{\neutf}{{\tilde{\chi}}_4^0}
\newcommand{\mneutf}{m_{{\tilde{\chi}}_4^0}}
\newcommand{\chargj}{{\tilde{\chi}}_j^+}
\newcommand{\chargop}{{\tilde{\chi}}_1^+}
\newcommand{\chargopm}{{\tilde{\chi}}_1^\pm}
\newcommand{\chargomp}{{\tilde{\chi}}_1^\mp}
\newcommand{\mchargop}{m_{{\tilde{\chi}}_1^+}}
\newcommand{\mchargopm}{m_{{\tilde{\chi}}_1^\pm}}
\newcommand{\chargtp}{{\tilde{\chi}}_2^+}
\newcommand{\chargtpm}{{\tilde{\chi}}_2^\pm}
\newcommand{\mchargtp}{m_{{\tilde{\chi}}_2^+}}
\newcommand{\mchargtpm}{m_{{\tilde{\chi}}_2^\pm}}
\def\mchargo{m_{\tilde{\chi}_1^+}}
\def\mchargt{m_{\tilde{\chi}_2^+}}
\def\chargom{\tilde{\chi}_1^-}
\def\chargtm{\tilde{\chi}_2^-}

\newcommand{\Att}{\mathrm{A}_{\tau \tau}}
\newcommand{\dtb}{\frac{\delta t_\beta}{t_\beta}}
\newcommand{\dtbeta}{\delta t_\beta}
\newcommand{\drbar}{{\overline{\rm DR}}}
\newcommand{\mhs}{{MH}}
\newcommand{\atts}{{\mathrm{A}_{\tau \tau}}}
\def\mct{m_{\neuto} m_{\chargop} m_{\chargtp}}
\def\mot{m_{\neuto} m_{\chargop} m_{\neutt}}
\def\motcoct{m_{\neutt} m_{\chargop} m_{\chargtp}}
\def\SloopS{\texttt{SloopS}}
\def\micromegas{{\texttt{micrOMEGAs}}}
\def\Comphep{\texttt{CompHEP}}
\def\Calchep{\texttt{CalcHEP}}
\def\Lanhep{\texttt{LanHEP}}
\def\SuSpect{\texttt{SuSpect}}
\def\ROOT{\texttt{ROOT}}

\def\tgb{\tan \beta}
\newcommand{\tb}{t_\beta}


\newcommand{\stau}{\tilde{\tau}}


\newcommand{\Omegah}{\Omega_{\chi} h^{2}}
\newcommand{\sigmav}{\langle \sigma v \rangle}


\newcommand{\beqn}{\begin{eqnarray}}
\newcommand{\eeqn}{\end{eqnarray}}
\newcommand{\bea}{\begin{eqnarray}}
\newcommand{\ena}{\end{eqnarray}}
\newcommand{\ra}{\rightarrow}
\newcommand{\susy}{{{\cal SUSY}$\;$}}
\newcommand{\su}{$ SU(2) \times U(1)\,$}

\begin{titlepage}

\begin{center}

\vspace*{3cm}

{\Large {\bf One-loop corrections, uncertainties and
approximations in neutralino annihilations: Examples}}

\vspace{8mm}

{F. Boudjema${}^{1)}$, \large G. Drieu La Rochelle${}^{1),2)}$ and S.~Kulkarni${}^{3)}$ }\\

\vspace{4mm}

{\it 1) LAPTh$^\dagger$, Univ. de Savoie, CNRS, B.P.110,
Annecy-le-Vieux F-74941, France}
\\ {\it
2) CERN Physics Department, Theory Division, CH-1211 Geneva 23, Switzerland}\\
 {\it
3) Bethe Center for Theoretical Physics and Physikalisches
Institut, Universit\"at Bonn, Nussallee 12,
D-53115 Bonn, Germany}\\

\vspace{10mm}

\today
\end{center}

\centerline{ {\bf Abstract} } \baselineskip=14pt \noindent

{\small The extracted value of the relic density has reached the
few per-cent level precision. One can therefore no longer content
oneself with calculations of this observable where the
annihilation processes are computed at tree-level, especially in
supersymmetry where radiative corrections are usually large.
Implementing full one-loop corrections to all annihilation
processes that would be needed in a scan over parameters is a
daunting task. On the other hand one may ask whether the bulk of
the corrections are taken into account through effective couplings
of the neutralino that improve the tree-level calculation and
would be easy to implement. We address this issue by concentrating
in this first study on the neutralino coupling to i) fermions and
sfermions and ii) $Z$. After constructing the effective couplings
we compare their efficiency compared to the full one-loop
calculation and comment on the failures and success of the
approach. As a bonus we point out that large non decoupling
effects of heavy sfermions could in principle be measured in the
annihilation process, a point of interest in view of the latest
limit on the squark masses from the LHC. We also comment on the
scheme dependencies of the one-loop corrected results.}

\vspace*{\fill}

\vspace*{0.1cm} \rightline{LAPTh-031/11}

\vspace*{1cm}

$^\dagger${\small UMR 5108 du CNRS, associ\'ee  \`a l'Universit\'e
de Savoie.} \normalsize

\vspace*{2cm}

\end{titlepage}

\renewcommand{\topfraction}{0.85}
\renewcommand{\textfraction}{0.1}
\renewcommand{\floatpagefraction}{0.75}

\section{Introduction}
With barely $1 \rm{fb}^{-1}$ of data, the LHC is pushing many
hitherto popular, though naive, extensions of supersymmetry to the
corners of high masses\cite{lhc_susy_limits_june2011} while
leaving some hope for a discovery of a rather light Higgs that
could still be compatible with
supersymmetry\cite{lhc_higgs_limits_june2011}. Before this very
recent paradigm, suspersymmetric models (and most models of new
physics for that {\em matter}) were very strenuously constrained
to a thin sliver in parameter space, most notably from the very
precise measurement of the dark matter relic density that has
now reached the few percent level and that will get even more
precise in the future, hence cornering even further model
building. Combining the results of  the 7-year {\tt WMAP} data
\cite{Jarosik:2010iu} on the 6-parameter $\Lambda$CDM model, the
baryon acoustic oscillations from {\tt SDSS}\cite{Percival:2009xn}
and the most recent determination of the Hubble
constant\cite{Riess:2009pu} one\cite{Komatsu:2010fb} arrives at
$\Omega_{{\rm CDM}}h^2 = 0.1123 \pm 0.0035$, where $\Omega_{{\rm
CDM}}$
 is the density of cold dark matter normalised to the critical density, and $h$ is the Hubble
constant in units of $100$ km s$^{-1}$Mpc$^{-1}$. One has reached a
precision of 3\%. The data from the LHC does not infer that dark
matter within supersymmetry, exemplified most nicely through the
neutralino, the lightest supersymmetric particle (LSP), is of
order of 1TeV or so, this is just a limit on the coloured
constituents of the model. As for the Higgs, were it not for
the large radiative corrections on the mass of the lightest
member, supersymmetry would long be a forlorn construct. The Higgs
is  the most prominent example where radiative corrections are far
from negligible in supersymmetry, yet practically all analyses
that aim at constraining the parameter space of the MSSM through
the relic density are based on tree-level cross sections of the
annihilation processes entering the predictions of this quantity
which as stressed is experimentally given within the percent
accuracy. Only seldom do some analyses assign a theory uncertainty
to these annihilation cross sections, an uncertainty due
essentially to the fact that higher order loop effects are not
known. This uncertainty, in the rare case where it is taken into account, is however assumed to be invariably the
same whatever the nature of the dominant process and the
composition of the LSP. The reason the loop corrections are
ignored, irrespective of the model specified, is that the calculation of the relic density requires most
often the evaluation of a
large number of processes. Most analyses are done with public
codes\cite{micromegas,darksusy,superiso-relic} based on tree-level
calculations. Computations of the relic density at one-loop have
now been achieved for quite a few
channels\cite{baro07,Freitas-relic-qcd,Klasen-relic-qcd,boudjema_chalons1,Bjorn_review_rc}
and tools exist now to perform in principle any calculation of the
relic density beyond tree-level amplitudes thanks to the recent
development of adapted automation tools\cite{boudjema_gondolo}.
Beside the
findings\cite{baro07,Freitas-relic-qcd,Klasen-relic-qcd,boudjema_chalons1,Bjorn_review_rc}
that these corrections are important, the improvements have not
percolated to most analyses. It must be said that these
calculations do involve some non trivial issues about the
renormalisation of the MSSM and more generally techniques for
one-loop integrals that certainly require expertise. The other
reason is that even when they could be implemented, they are still
certainly extremely CPU time consuming, forbidding hence any
attempts of fits, likelihood search, and in a more general context
any sampling of the parameter space, especially if one takes into
account the fact that the MSSM is more than liberal with
unconstrained parameters. Yet, apart from providing a more precise
prediction, one-loop calculations can probe higher masses, a
situation akin to the precision electroweak observables and their
sensitivity to the top and Higgs mass. For example, non decoupling
effects termed in analogy with electroweak SM observables,
super-oblique corrections have been revealed in one-loop
calculations of supersymmetry
observables\cite{feng_nondecoupling,randall_nondecoupling,nojiri_nondecoupling}.
An example in view of the recent findings of the LHC is that super
heavy squarks leave a non negligible imprint on many observables
in particular the annihilation cross sections
involving dark matter. \\
The aim of this paper is two-fold. First, to stress again the
importance of the loop corrections for the relic density and show
again that even when a loop calculation is available, there still
remains in some cases an uncertainty that pertains to the choice
of the renormalisation scheme. The second and more detailed aspect is to discuss
whether an approximation to the one-loop calculation can be found.
We aim at implementing a universal correction through effective
couplings of the LSP and check its validity against a complete
one-loop calculation. If such an approximation is possible and
general enough it could be implemented in existing codes
(based on tree-level cross sections) calculating the relic
density. Such was the case with the inclusion of the Sommerfeld
effect\cite{Sommerfeldinclude} in the case of coannihilation or
processes dominated by Higgs exchange for which $\Delta
m_b$\cite{delta_mb} corrections are included\cite{micromegas}.

This preliminary study takes a simple process, namely $\neuto
\neuto \ra \mu^+ \mu^-$, as a testing ground. The aim of this
study is not to find a good scenario that returns the correct
actual value of the relic density but to try to unravel some
general common features of the loop calculation to improve the
predictions of the relic density. The aim is to rather find out
whether one can improve on the tree-level calculation by
introducing effective couplings of the LSP that could be used for
any process. As we will see, though at first sight naive, the
process $\neuto \neuto \ra \mu^+ \mu^-$ embodies the three types
of  couplings of the neutralino: to $f/\tilde{f}$, gauge bosons
and Higgses. Here we
cover the bino and the higgsino case. 
One might argue that the bino case corresponds to what was
referred to as the bulk region in the constrained MSSM and is
largely ruled out, whereas for a higgsino this would not be the
dominant process. As we have just stressed the aim is not to
strive to find a good scenario. Besides, it suffices to change the
cosmological ingredients\cite{nonconventional-relic} entering the
calculation of the relic density to revive the so called bulk
region. We will see that by considering these few simple examples
the conclusions about the efficacy of the effective coupling is
quite different. Moreover, there is no need here to convert a full
corrected cross section into a relic density value, we rather take
for all the models we study the annihilation cross section for an
energy corresponding to a relative velocity of $v=0.2$, typical
for the relic density calculation. As known, for zero relative
velocity the process enjoys chirality suppression which is lifted
at higher order through gauge boson radiation, however the effect
on the relic is totally marginal\cite{baro07}.

The paper is organised as follows. We first briefly describe the
ingredients necessary to perform a one-loop calculation in
supersymmetry covering both automation, renormalisation and
renormalisation schemes, it is in this section that we will write
down the effective   universal neutralino couplings as well as
some definitions. Section~3 contains our main results. After a few
definitions we first study the case of a bino-like neutralino
before addressing the case of a higgsino-like LSP. We also
quantify the possible uncertainties due to the scheme
dependence. We conclude in Section~4 by some general observations.\\
Throughout the paper we use some shorthand notation for angles.
Generically $c_\theta, s_\theta, t_\theta$ stand  for $\cos
\theta, \sin \theta, \tan \theta$. The weak mixing angle
$\theta_W$ is defined as $\cos \theta_W=c_W=M_W/M_Z$ where $M_W$
is the $W$ mass and $M_Z$ the $Z$ mass.

\section{Calculations, renormalisation, schemes at full one-loop}

\subsection{Tree-level considerations}
\begin{figure*}[htbp]
\begin{center}
\begin{tabular}{ccc}
\includegraphics[width=0.3\textwidth,height=0.2\textwidth]{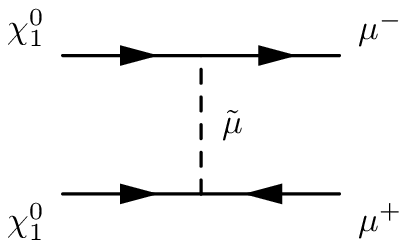}&
\includegraphics[width=0.3\textwidth,height=0.2\textwidth]{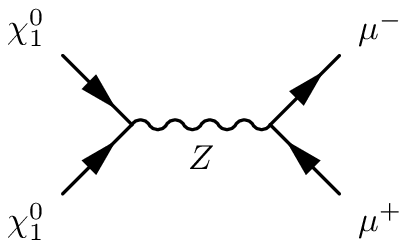}&
\includegraphics[width=0.3\textwidth,height=0.2\textwidth]{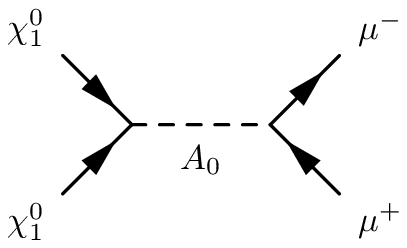}
\\
&& \\
$(a)$ & $(b)$ & $(c)$ \\
& &
\end{tabular}
\end{center}
\caption{\label{fig:tree_diags_neut1neut1mumu}{\em  Tree-level
diagrams contributing to $\neuto \neuto \ra \mu^+ \mu^-$. ($a$) is
the $t$-channel $\tilde{\mu}$ exchange, ($b$) is the $Z$ exchange
and ($c$) an example of a Higgs exchange.}}
\end{figure*}

At tree-level, see Fig.~\ref{fig:tree_diags_neut1neut1mumu},
$\neuto \neuto \to \mu^+ \mu^-$ proceeds through {\it i})
$t$-channel smuon exchange dominated by a $\tilde{\mu}_R$ in the
case of a bino-like since it has the largest hypercharge, {\it
ii}) a $Z$ exchange which, on the other hand, is suppressed for
the bino {\it iii}) Higgs exchange but this is small in view of
the Yukawa coupling of the muon. Therefore, as advertised, all types
of couplings for the LSP are present: to fermions in the $\neuto
\tilde{\mu} \mu$ coupling, gauge bosons in the $\neuto \neuto Z$,
and Higgs scalars such as $\neuto \neuto A^0$ ($A^0$ is the
pseudoscalar Higgs). It is through the choice of a hierarchy in
the set $M_1,M_2,\mu$ that we can largely define the nature of the
LSP. Numericaly speaking we call a neutralino pure or almost pure
when its mixing to the specified species is over 99\%.

\subsection{Renormalisation and loop corrections, general considerations and issues}

\subsubsection{Set up of the automatic calculation:
\texttt{SloopS}} One-loop processes calculated via the
diagrammatic Feynman approach involve a huge number of diagrams
even for $2 \rightarrow 2$ reactions, especially in a theory like
supersymmetry. Performing a  full one-loop calculation by hand
without automation is practically untractable. Our exact full
one-loop calculation is done with the help of the automated code
\texttt{SloopS}. \texttt{SloopS} is an automated code for one-loop
calculations in supersymmetry. It is a combination of
\texttt{LanHEP}\cite{lanhep}, the bundle
\texttt{FeynArts}\cite{feynarts}, \texttt{FormCalc}\cite{formcalc}
and an adapted version of \texttt{LoopTools}\cite{looptools,
boudjema_temes}. \texttt{LanHEP} deals with one of the main difficulties
that has to be tackled for the automation of the implementation of
the model file, which is entering the thousands
of vertices that define the Feynman rules. On the theory side a
proper renormalisation scheme needs to be set up, which then means
extending many of these rules to include counter-terms. This part
is done through \texttt{LanHEP} which allows to shift fields and
parameters and thus generates counterterms most efficiently. The
ghost Lagrangian is derived directly from the BRST
transformations. The loop libraries used in \texttt{SloopS} are
based on \texttt{LoopTools} with the addition of quite a few
routines in particular those for dealing with small Gram
determinants that appear in our case at small relative velocities
of the annihilating dark matter, and even more so of relevance for
indirect detection\cite{boudjema_temes}.

\subsubsection{Renormalisation}
In \texttt{SloopS} all sectors of
the MSSM are implemented through a one-loop  renormalisation. This
is explained in details
in\cite{baro07,Sloops-higgspaper,baro09,boudjema_chalons1}. Here
we only briefly sketch the renormalisation procedure. We have
worked, as far as possible, within an on-shell scheme generalising
what is done for the
electroweak standard model\cite{grace-1loop}.\\

\noi {\it {\bf i)}} The Standard Model parameters : the fermion masses
as well as the mass of the W and Z are taken as input physical
parameters. The electric charge is defined in the Thomson limit,
see for example\cite{grace-1loop}. The light quarks (effective)
masses are chosen \cite{eennhletter} such as to reproduce the SM
value of $\alpha^{-1}(M_Z^2)$ = 127.77. This should be kept in
mind since one would be tempted to use a $\drbar$ scheme for
$\alpha$, defined as $M_Z$, to take into account the fact that
dark matter is annihilating at roughly the electroweak scale, so
that $\alpha((2  \mneuto)^2)$ is a more
appropriate choice. One should remember that
the use of $\alpha((2 \mneuto)^2)$ instead of the on-shell value
in the Thomson limit would correct the tree-level cross section
for $\neuto \neuto \ra \mu^+ \mu^-$ by about $14\%$.  As we will
see and have reported somewhere else for other processes this running
does not, most of the time, take into account the bulk of the
radiative corrections that we report here. Therefore for further
reference, let us introduce the correction due to the running of
$\alpha$,
\beqn
\label{eq:running_alpha}
 \Delta_{\alpha}=\frac{\saeff-\s0}{\s0}=2 \Delta \alpha,
\eeqn
where the cross section $\s0$ is the tree-level calculated with
$\alpha_0=\alpha(0)=1/137.0359895$ whereas $\saeff$ is the tree-level with
$\alpha_0 \ra \alpha_{\rm eff.}(Q^2)=\alpha(Q^2)=\alpha_0/(1+\Delta
\alpha(Q^2)), Q=2 \mneuto$. With our input parameters $\Delta
\alpha(M_Z^2)=0.06$. In the running we allow for all charged
particles including the $W$ boson contribution, the top and the
sfermions and the charginos, though for the light LSP scenario we
consider these added contributions are very small\footnote{For the
$W$ boson contribution the self energy of the photon is calculated
in a non-linear gauge\cite{grace-1loop} corresponding to the
background gauge in order to maintain $U(1)_{\rm QED}$ gauge
invariance.}

\noi {\it {\bf ii)}} The Higgs sector : The pseudoscalar Higgs
 mass $M_A^0$ is used as an input parameter while insisting on vanishing tadpoles.
$\tb$, which at tree-level is the ratio of the two expectation
values of the Higgs doublets, can be defined through several
schemes:
\begin{description}
    \item[~-] In the DCPR scheme\cite{DabelsteinHiggs,DCPR} $\tb$ is defined by requiring that the
(renormalised) $A^{0}Z$ transition vanishes at
$Q^2=M_{A^0}^2$.
       \item[~-] a $\drbar$ definition where the $\tb$ counter-term, $\delta t_\beta$, is defined as a
       pure divergence leaving out all finite parts.
    \item[~-] a process-dependent definition of this counter-term by extracting it
    from the decay $A^0 \rightarrow \tau^+\tau^-$ that we will refer to as $\atts$ for
    short. This definition is a good choice for the gauge independence of the processes.
    \item[~-] an on-shell definition with the help of the mass of the heavy CP Higgs
    $H$ taken as input parameter called the MH scheme from now on.
    We have reported elsewhere that this scheme usually
    introduces large radiative corrections.

\end{description}
These schemes are critically reviewed in \cite{Sloops-higgspaper}.
By default we use the DCPR scheme but when  quantifying the effect
of the scheme dependence on $\tb$ we also use the $\drbar$ and MH
scheme.

\noi {\it {\bf iii)}} The sfermion sector : For the process at hand only the smuons parameters
require renormalisation.  For the slepton sector we use as input
parameters masses of the two charged sleptons which in the case of
no-mixing define the R-slepton soft breaking mass, $M_{{\tilde
\mu}_{R}}$ and the $SU(2)$ mass,
 $M_{{\tilde \mu}_{L}}$, giving a correction to the sneutrino mass at one-loop.
Though not needed here, in the squark sector each generation needs
three physical masses to constrain the breaking parameter
$M_{{\tilde Q}_{L}}$ for the $SU(2)$ part, $M_{{\tilde u}_{R}}$,
$M_{{\tilde d}_{R}}$ for the R-part. See\cite{baro09} for details.\\

\noi {\it {\bf iv)}} The chargino/neutralino sector.
First of all, for the neutralinos at tree-level the physical
fields $\chi^0_i,i=1,\dots,4$ are obtained from the current
eigenstates $\big(\psi^{n}\big)^t=(\tilde{B}^0,\tilde{W}^{0},
\tilde{H}_1^0,\tilde{H}_2^0)$ through a unitary complex matrix $N$
\begin{eqnarray}
\label{eq:N-neut} \chi^{0}=N \psi^{n}.
\end{eqnarray}
$N$ diagonalises the mass mixing matrix $Y$ in the neutralino
sector, see~\cite{baro09} for details and conventions. Although
only $\neuto$ enters our calculations we do need to fix all the
elements that define its composition and hence couplings. For this
sector we implement an on-shell scheme by taking as input three
masses in order to reconstruct the underlying parameters
$M_1,M_2,\mu$. In \SloopS~\cite{baro09} the default scheme is to
choose two charginos masses $\mchargopm$ and $\mchargtpm$ as input
to define $M_2$ and $\mu$ and one neutralino mass to fix $M_1$.
The masses of the remaining three neutralinos receive one-loop
quantum corrections. In this scheme, these counterterms
are~\cite{baro09}
\begin{eqnarray}
\delta M_{2}&=&\frac{1}{M_{2}^{2}-\mu^{2}}\left(
(M_{2}\mchargo^2-\mu {\rm det} X)\frac{\delta\mchargo}{\mchargo}
+ (M_{2}\mchargt^2-\mu {\rm det} X)\frac{\delta\mchargt}{\mchargt}\right.\nonumber\\
& &-\left. M_{W}^2(M_{2}+\mu s_{2\beta})\frac{\delta
M_{W}^2}{M_{W}^2} - \mu M_{W}^{2} s_{2 \beta}c_{2 \beta}
 \frac{\delta t_\beta}
{t_{\beta}}\right),\nonumber\\
\delta \mu &=&\frac{1}{\mu^{2}-M_{2}^{2}}\left(
(\mu\mchargo^2-M_{2} {\rm det} X)\frac{\delta\mchargo}{\mchargo}
+ (\mu\mchargt^2- M_{2}{\rm det} X)\frac{\delta\mchargt}{\mchargt}\right.\nonumber\\
& &-\left. M_{W}^2(\mu+M_{2} s_{2\beta})\frac{\delta
M_{W}^2}{M_{W}^2} - M_{2} M_{W}^{2} s_{2 \beta} c_{2 \beta}
\frac{\delta t_\beta}
{t_{\beta}}\right),\\
\delta M_{1}&=&\frac{1}{N_{1i}^{2}}(\delta
m_{\chi_{i}^{0}}-N_{2i}^{2}\delta
M_{2}+2N_{3i}N_{4i}\delta\mu\nonumber
\\& &-2N_{1i}N_{3i}\delta
Y_{13} - 2N_{2i}N_{3i}\delta Y_{23} - 2N_{1i}N_{4i}\delta Y_{14} -
2N_{2i}N_{4i}\delta Y_{24} )\, , \label{dm1dm2dmu}
\end{eqnarray}
with $\textrm{det}X=M_{2}\mu-M_{W}^{2}s_{2\beta}$~\cite{baro09}.
$\delta m_{\tilde{\chi}_{i}^{0}}$ is the counterterm of $i$th neutralino
defined entirely from its self-energy,
see\cite{baro09}. $\delta O$ represents the shift on the parameter $O$
that generates the counterterm for that quantity. \\
Looking at these equations some remarks can be made. First, in the
special configuration $M_2 \sim \pm\mu$ an apparent singularity
might arise. Ref.~\cite{baro07} pinpointed this configuration
which can induce a large $\tb$-scheme dependence in the
counterterms $\delta M_{1,2}$ and $\delta\mu$.
Such mixed scenario is not covered here.  \\
Second, the choice of $\mneuto$ as an input parameter is
appropriate only if the lightest neutralino is mostly bino
($|N_{11}| \sim 1$) or if the bino like neutralino is not too
heavy compared to other neutralinos. Indeed we can see that if
$N_{1i} \sim 0$ the counterterm $\delta M_1$ is subject to large
uncertainty and may  introduce large finite correction, this is
related to the fact that $M_1$ is badly reconstructed. To avoid
such uncertainty we only choose $i$ as the most bino like, in
other words in Eq.~\ref{dm1dm2dmu}, $|N_{1i}|=Max(|N_{1j}|),
j=1,..4$.

\noi {\it {\bf v)}} Finally diagonal field renormalisation is fixed by demanding
that the residue at the pole of the propagator of all physical
particles to be unity, and the non-diagonal part by demanding
no-mixing between the different physical particles when on shell.
This is implemented in all the sectors. In our case apart from the
muon, this step is important for the $\neuto$. We insist that
$N_{ij}$ is used both at the tree-level and one-loop level.
Nonetheless to define the physical state we do introduce the shift
for the neutralinos\cite{baro09} through wave function
renormalisation

\begin{eqnarray}
\label{eq:wfrneut} \tilde{\chi}_{i}^{0} \to
\tilde{\chi}_{i}^{0}+\frac{1}{2}\sum_j \left(\delta
Z_{ij}P_{L}+\delta Z^{*}_{ij}P_{R}\right)\tilde{\chi}_{j}^{0} \, .
\end{eqnarray}

\noi {\it {\bf vi)}} Dimensional reduction is used as implemented in the
\texttt{FFL} bundle at one-loop through the equivalent constrained
dimensional renormalisation\cite{CDR}.

\subsubsection{Infrared divergences} For the processes $\neuto\neuto
\rightarrow \mu^+ \mu^- $,  we can decompose the one-loop
amplitudes in a virtual part $\mathcal{M}_{1loop}^{EW}$ and a
counter-term contribution $\mathcal{M}_{CT}$. The sum of these two
amplitudes must be ultraviolet finite and gauge independent. Due
to the virtual exchange of the massless photon, this sum can
contain infrared divergencies. This is cured by adding a small
mass to the photon and/or gluon, $\lambda_{\gamma}$ and
$\lambda_g$.  This mass regulator should exactly cancel against
the one present in the final state radiation of a photon. The QED
contribution is therefore split into two parts : a soft one where
the photon energy $E_{\gamma}$ is integrated to less than some
small cut-off $k_c$ and a hard part with $E_{\gamma} > k_c$. The
former requires a photon mass regulator. Finally the sum
$\mathcal{M}_{1loop}+\mathcal{M}_{CT}+\mathcal{M}_{\gamma}^{soft}(E_{\gamma}
< k_c)+ \mathcal{M}_{\gamma,g}^{hard}(E_{\gamma,g} > k_c)$ should
be ultraviolet finite, gauge invariant, not depend on the mass
regulator and on the cut $k_c$.

\subsubsection{Checking the result}
\noi {\it {\bf i)}} For each process and set of parameters, we
first check the ultraviolet finiteness of the results. This test
applies to the whole set of virtual one-loop diagrams. The
ultraviolet finiteness test is performed by varying the
ultraviolet parameter $C_{UV}=1/\varepsilon$, $\varepsilon$ is the
usual regulator in dimensional reduction. We vary $C_{UV}$ by
seven orders of magnitude with no change in the result. We content
ourselves with double precision.\\

\noi {\it {\bf ii)}} The test on the infrared finiteness is
performed by including both the loop and the soft bremsstrahlung
contributions and checking that there is no dependence on the
fictitious photon mass $\lambda_{\gamma}$.\\

\noi {\it {\bf iii)}} Gauge parameter independence of the results
is essential. It is performed through a set of the {\em eight}
gauge fixing parameters based on the implementation of a
non-linear gauge\cite{Sloops-higgspaper}.

\subsection{Effective couplings for neutralino
interactions {\it vs} Full calculation}

\subsubsection{Contributions at full one-loop}
\begin{figure*}[htbp]
\begin{center}
\begin{tabular}{ccc}
\includegraphics[width=0.3\textwidth,height=0.2\textwidth]{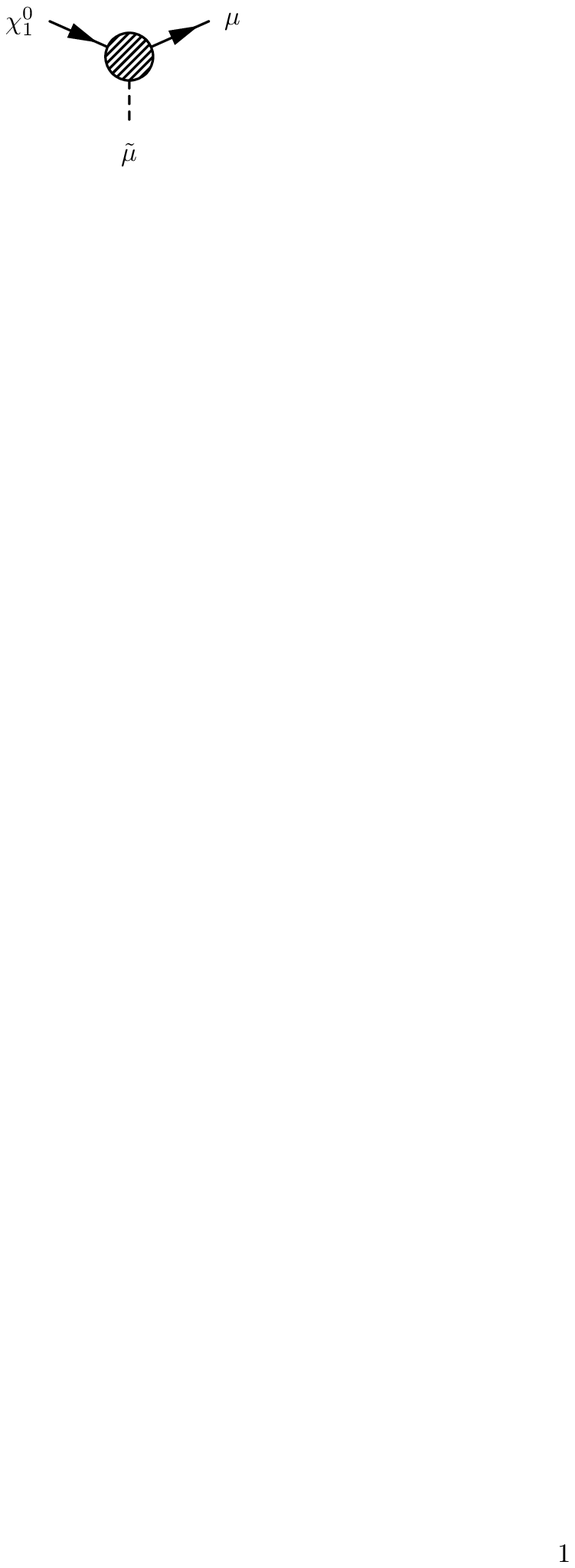}&
\includegraphics[width=0.3\textwidth,height=0.2\textwidth]{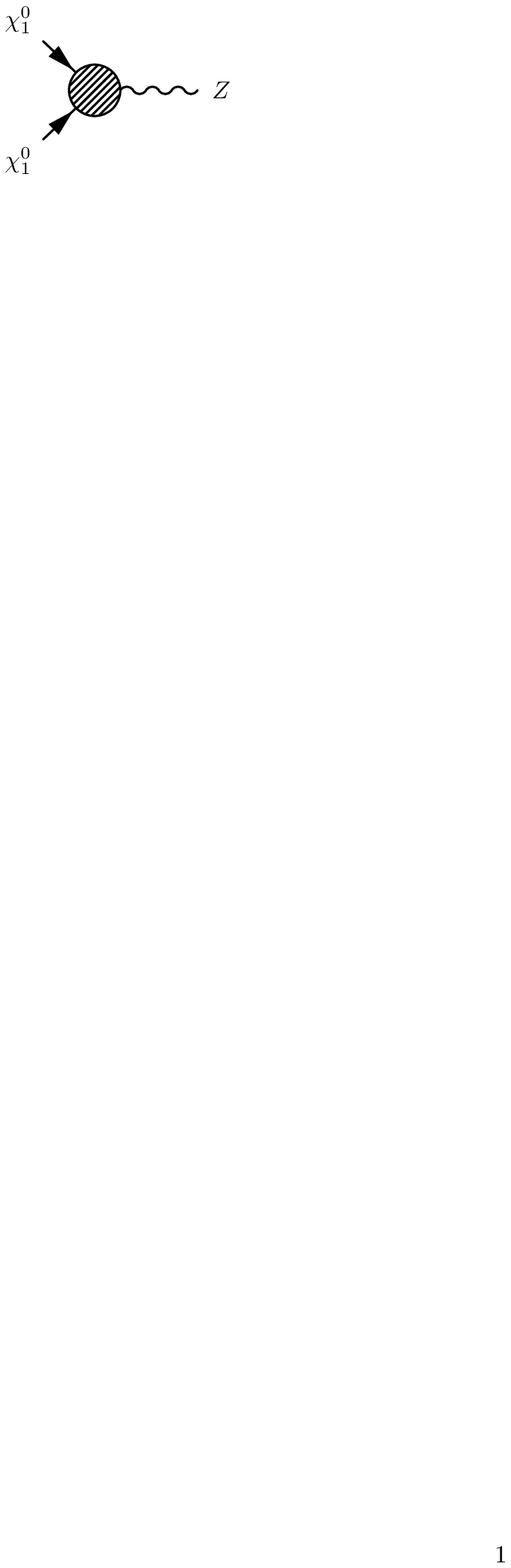}&
\includegraphics[width=0.3\textwidth,height=0.2\textwidth]{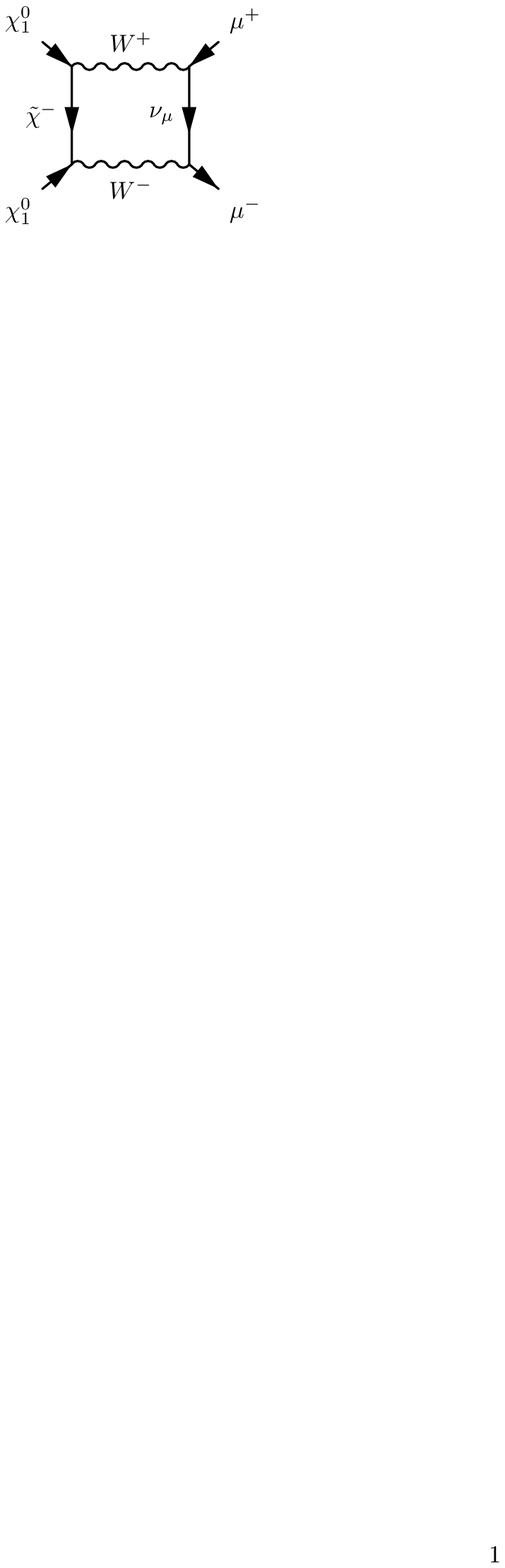}
\\
&& \\
(a) & (b) & (c) \\
& &
\end{tabular}
\end{center}
\caption{\label{fig:loop_diags_neut1neut1mumu}{\em  Different
types of corrections appearing at one-loop for the process $\neuto
\neuto \ra \mu^+ \mu^-$. ($a$) is the correction to vertex $\neuto
\mu \tilde{\mu}$ for the $t$-channel $\tilde{\mu}$ exchange, ($b$)
is the full correction to the $\neuto \neuto Z$ vertex for the
$s$-channel $Z$ exchange and ($c$) is an example of a box loop.}}
\end{figure*}
The full set of one-loop contributions to the process $\neuto
\neuto \ra \mu^+ \mu^-$ consist of two-point functions
(self-energies and transitions such as $\neuto \ra \neutt$),
vertex three-point functions as in
Figs~\ref{fig:loop_diags_neut1neut1mumu}(a,b) and box diagrams.
The vertex corrections include also the counterterms, the latter
as explained previously involve two-point functions. To these, one
should also add the QED final state radiation.

\subsubsection{Effective couplings of the neutralino at one-loop}
Among this full set of corrections one can construct a
finite subset that is not specific to the muon. This subset will
be involved in all processes involving neutralinos. For example,
the vertex correction to $\neuto \neuto Z$ is obviously
independent of the muon being in the final state, a similar
statement can be said for $\neuto \neuto h/H/A$. Also, all
occurrences of the wave function renormalisation of the neutralino
(including transitions between neutralinos) and the $Z$ are
process independent. The same can be said  also of the
counterterms to the gauge couplings and the vacuum expectation
values or in other words $v,\tb$. On the other hand the wave
function renormalisation of the muon is specific to the muon final
state. The boxes the four-point one particle irreducible (1-PI)
functions, as well as the QED correction are also specific to the
process. The construct of the universal correction for the
effective coupling $\neuto f \tilde{f}$ from $\neuto \mu
\tilde{\mu}$ is different from that of $\neuto \neuto Z$, since in
the latter all three particles can be considered as universal. For
example the full correction to the vertex $\neuto \mu \tilde{\mu}$
shown in Fig.~\ref{fig:loop_diags_neut1neut1mumu}(a), consists of
a 1-PI 3-point function vertex correction (triangle) which is
muon specific and that does not need to be calculated to build up
the effective coupling. It also contains wave function
renormalisation of the neutralinos as well as  counterterms for
the gauge couplings and for other universal quantities such as
$\tb$ which must be combined to arrive at the universal
correction for the $\neuto f \tilde{f}$ vertex. The aim of the
paper is therefore to extract these process independent
contributions and define effective vertices for the LSP
interactions. This is akin to the effective coupling of the $Z$ to
fermions where universal corrections are defined. Describing the
bulk of the radiative corrections in terms of effective couplings
has been quite successful to describe for example the observables
at the $Z$ peak. Although not describing most perfectly the effect
of the full corrections for all observables (for example $Z b \bar
b$ receives  an important triangle contribution due to the large
top Yukawa coupling) one must admit that the approach has done quite a
good job. Most of the effective corrections were universal,
described in terms of a small set of two-point functions of the
gauge bosons. \\
The other benefit was that such approximations were sensitive to
non decoupling effects that probe higher scales (top mass and Higgs mass). The
set of two-point functions, and for $\neuto \neuto Z$ three-point
functions, should of course lead to a finite and gauge invariant
quantity. Loops involving gauge bosons have always been problematic (even in
the case of the $Z f \bar{f}$) in such an approach
since it is difficult to extract a gauge independent value. For
the couplings of the neutralinos as would be needed for
approximating their annihilation cross section independently of
the final state, one would therefore expect that apart from the
rescaling of the gauge couplings which can be considered as an
overall constant, the mixing effect between the different
neutralinos should be affected. One can in fact re-organise a few
of the two point functions (that can be written also as
counterterms) to define an effective coupling for the neutralino.
One should of course also correct in this manner the $Z \mu^+
\mu^-$ coupling. Let us stress again that in this first
investigation we will primarily take into account the effects of
fermions and sfermions in the universal loops. For the $\neuto
\neuto Z$ effective we also attempt to include the virtual
contribution of the gauge bosons especially that for the
higgsino-like the coupling to $W$ and $Z$ are not suppressed.

\subsubsection{The effective $\neuto f \tilde{f}$}
To find the process independent corrections to this coupling, we
recall that in the basis $(\tilde{B}^0,\tilde{W}^{0},
\tilde{H}_1^0,\tilde{H}_2^0)$ before mixing and for both $f_{L,R}$
the couplings for the two chiral Lorentz structures writes as
\beqn
\label{eq:wbh_fsf} \frac{1}{\sqrt{2}} \big( g^\prime Y_f, g
\tau^3_f, y_{1,f}, y_{2,f} \big)=\frac{1}{\sqrt{2}} \big( g^\prime
Y_f, g \tau^3_f, \frac{g m_u}{M_W c_\beta},  \frac{g m_d}{M_W
s_\beta}\big) \rightarrow \big( g^\prime , g , \frac{g }{M_W
c_\beta}, \frac{g }{M_W s_\beta}\big),
\eeqn

$Y_f,\tau_3^f$ are the isospin and $SU(2)$ charges of the
corresponding fermion/sfermions. The two higgsinos couple
differently to the up and down fermions with a coupling that is
proportional to the Yukawa coupling. Though  this is not universal
we can still isolate a universal part where there is no  reference
to the final fermion/sfermion. This is what is meant by the last
expression in Eq.~\ref{eq:wbh_fsf} where the explicit mass of the
corresponding fermion masses has been dropped. The
variations/counterterms on these parameters have to be implemented
before turning to the physical basis. The latter as explained in
the previous paragraph is achieved through the diagonalising
matrix $N$ (Eq.~\ref{eq:N-neut}) as in tree-level supplemented by
wave function renormalisation which involve both diagonal and non
diagonal transitions of the neutralino, see Eq.~\ref{eq:wfrneut}.
In the case of effective coupling of neutralinos,  this is
achieved by defining an effective mixing matrix such that $N \ra
N+\Delta N^{\chi f \tilde{f}}$ in all couplings of the neutralino.
The $\Delta N^{\chi f \tilde{f}}$ write as

\beqn
\label{eq:app_hollik}
\Delta N_{i1}^{\chi f \tilde{f}}&=&\frac{\delta g^\prime}{g^\prime} N_{i1} + \frac{1}{2}\sum_{j}N_{j1}\delta Z_{j i}, \nonumber \\
\Delta N_{i2}^{\chi f \tilde{f}}&=&\frac{\delta g}{g}N_{i2} + \frac{1}{2} \sum_{j}N_{j2}\delta Z_{j i}, \nonumber \\
\Delta N_{i3}^{\chi f \tilde{f}}&=&\left(\frac{\delta g}{g}-\1\frac{\delta M_W^2}{M_W^2}-\frac{\delta c_{\beta}}{c_{\beta}}\right)N_{i3}+\frac{1}{2}\sum_{j}N_{j3}\delta Z_{j i}, \nonumber \\
\Delta N_{i4}^{\chi f \tilde{f}}&=&\left(\frac{\delta
g}{g}-\1\frac{\delta M_W^2}{M_W^2}-\frac{\delta
s_{\beta}}{s_{\beta}} \right)N_{i4} + \frac{1}{2}
\sum_{j}N_{j4}\delta Z_{j i}.
\eeqn
where $j$ runs from 1 to 4 and for the LSP , $i=1$.

All the counterterms above are calculated from self-energy
two-point functions and are fully defined in
~\cite{Sloops-higgspaper,baro09}. $\delta g/g=\delta e/e-\delta
s_W/s_W,\delta g^{\prime}/g^{\prime}=\delta e/e-\delta c_W/c_W$.
$\delta s_\beta/s_\beta=c_\beta^2 \delta \tb/\tb$.
Eq.~\ref{eq:app_hollik} agrees with what was suggested in
\cite{Hollik_susyeff}. Let us stress again that in these
self-energies no gauge bosons and therefore no neutralinos and
charginos are taken into account but just sfermions and fermions, otherwise this would not be finite.
For a bino-like, self-energies containing gauge and Higgs bosons
(with their supersymmetric conterparts) are not expected to
contribute much. This is not necessarily the case for winos and
higgsinos.

\subsubsection{The effective $\neuto \neuto Z$}
Since all particles making this vertex can now be considered as
being process independent (as far as neutralino annihilations are
concerned), all counterterms including wave function
renormalisation of both the $Z$ and $\neuto$ must be considered.
The price to pay now is that the genuine triangle vertex
corrections  $\neuto \neuto Z$ must also be included. It is only
the sum of the vertex and the self-energies that renders a finite
result. When correcting this vertex one must also correct the $Z
\mu^+ \mu^-$ vertex keeping within the spirit of calculating the
universal corrections. This can be implemented solely through
self-energy corrections (excluding the muon self-energies) and
there is no need to calculate here the genuine vertex corrections.
An exception would be the production of the $b$ and to some extent
the top where genuine vertex corrections are important. Talking of
heavy flavours, when computing the correction to
the $\neuto \neuto Z$ with the Z off shell with an invariant mass $Q^2$, one should also include the $\neuto \neuto
G^0$ vertex, where $G^0$ is the neutral Goldstone boson. In our
case we restrict ourselves to almost massless fermions. The case
of the top and bottom final states will be addressed elsewhere
together with the potential relevant
contribution of the Higgses in the $s$-channel. \\
Since one is including the genuine 1-PI vertex correction, it is
important to inquire whether this correction generates a new
Lorentz structure beyond the one found at tree-level. The
contribution to the tree-level Lorentz structure is finite after
adding the self-energies and the vertex. Any new Lorentz structure
will on the other hand be finite on its own. General arguments
based on the Majorana nature of the neutralinos backed by our
numerical studies show that no new Lorentz structure is generated
for neutralinos. First of all, at tree-level one has only one
structure
\beqn
{\cal {L}}_{\neuto \neuto Z}=\frac{g_Z}{4}
\Big(N_{13}N_{13}-N_{14} N_{14} \Big) \neuto \gamma_\mu
\gamma_5\neuto Z^\mu, \quad \quad g_Z=\frac{e}{c_W s_W}.
\eeqn
The overall strength is a consequence of the fact that the
coupling emerges solely from the higgsino with a gauge coupling.
Indeed in the $(\tilde{B}^0,\tilde{W}^{0},
\tilde{H}_1^0,\tilde{H}_2^0)$ basis the coupling is $\propto g_Z
(0,0,1,-1)$. Only the Lorentz structure $\gamma_\mu \gamma_5$
survives as a consequence of the Majorana nature. With $p_1,p_2$
denoting the incoming momenta of the two $\neuto$, at one-loop a
contribution $(p_1^\mu-p_2^\mu)$ does not survive symmetrisation,
whereas $(p_1^\mu+p_2^\mu)$ will not contribute for massless
muons. We calculate this correction  for a $Z$ with an invariant
mass $Q^2$, in the application this $Q^2$ will be set to the
invariant mass of the muon pair.  This vertex contribution is
denoted $\Delta g_{\neuto \neuto Z}^{\bigtriangleup}(Q^2) \equiv
\Delta g_{\neuto \neuto Z}^{\bigtriangleup}$. The contribution of
the coupling counterterms defining $g_Z$ \underline{and} the $Z$
wave function renormalisation define the universal correction to
the $Z$ coupling strength $g_Z^{\rm eff}=g_Z (1+\Delta g_Z)$,
with $\Delta g_Z/g_Z= \delta g_Z/g_Z+ \delta Z_{ZZ}/2$. $\delta
Z_{ZZ}$ is the wave function renormalisation of the $Z$. We of
course have to add the wave function renormalisation of the
$\neuto$ like what was done for the $\neuto f \tilde{f}$ vertex.
We improve on this implementation by taking into account the fact
that the $Z$ is off-shell and therefore the wave function
renormalisation through $\delta Z_{ZZ}=\Pi_{ZZ}^{\prime}(M_Z^2)$ is
only part of the correction that would emerge from the correction
to the complete $Z$ propagator in the $s$-channel contribution with invariant mass $Q^2$.
Note that here there is no need for including a $Z\gamma$
transition since photons do not couple to neutralinos. Collecting
all these contributions, the
effective vertex is obtained by making \\
$g_Z \to g_{\neuto \neuto Z}^{{\rm eff}}$ and $N_{i 1} \to N_{i 1}
+ \Delta N_{i 1}^{\neuto \neuto Z}$ with
\beqn
g_{\neuto \neuto Z}^{{\rm eff}}&=&g_Z (1+ \Delta g_Z(Q^2) + \Delta
g_{\neuto \neuto Z}^{\bigtriangleup}(Q^2)); \\ \quad \Delta
N_{ij}^{\neuto \neuto Z}&=&\frac{1}{2} \sum_{k} N_{kj} \delta
Z_{ki}, \;\; (i,j,k)=1\dots 4.
\eeqn
Explicitly
\def\sw2{s_W^2}
\def\cw2{c_W^2}
\def\MZ2{M_Z^2}
\def\MW2{M_W^2}
\beqn
\label{delgzuni} \Delta
g_Z&=&\1\left(\Pi'_{\gamma\gamma}(0)-2\frac{s_W}{c_W}\frac{\Pi_{\gamma
Z}(0)}{\MZ2}\right)+\1\left(1-\frac{\cw2}{\sw2}\right)
\left(\frac{\Pi_{ZZ}(\MZ2)}{\MZ2}-\frac{\Pi_{WW}(\MW2)}{\MW2}\right)
\nonumber \\
&-&\frac{1}{2} \left(
\frac{\Pi_{ZZ}(Q^2)-\Pi_{ZZ}(M_Z^2)}{Q^2-\MZ2} \right)\ .
\eeqn
At the same time for the fermion with charge $q_f$ we correct the
$Z f \bar f$ $\propto g_Z (\gamma_5 +(1-4 |q_f| s_W^2))\gamma_\mu$
by effectively making $g_Z \to g_Z (1+\Delta g_Z)$ with $\Delta
g_Z$ defined in Eq.~\ref{delgzuni} and $s_W^2$ to
\beqn
\label{delsweff} \Delta
\sw2=\frac{\cw2}{\sw2}\left(\frac{\Pi_{ZZ}(M_Z^2)}{\MZ2}-\frac{\Pi_{WW}(M_W^2)}{\MW2}\right)+
\frac{c_W}{s_W}\frac{\Pi_{\gamma Z}(k^2)}{k^2}\ .
\eeqn
By default we include only the fermions and sfermions in the
virtual corrections described by
Eqs.~\ref{delgzuni}-\ref{delsweff}. For the $\neuto \neuto Z$ one
expect the contribution of the gauge bosons and the
neutralinos/charginos to be non negligible especially for the
higgsino case. In fact, including such contributions still gives
an ultraviolet finite result for $g_{\neuto \neuto Z}^{{\rm eff}}$
in Eq.~\ref{delgzuni} which is a non trivial result. Moreover this
contribution is gauge parameter independent in the class of
(linear) and non-linear gauge fixing conditions\cite{grace-1loop}.
To weigh up the gauge/gaugino/higgsino contribution we will
therefore also compare with this generalised effective $g_{\neuto
\neuto Z}^{{\rm eff}}$ including all virtual particles. Observe that in Eq.~\ref{delgzuni} we have
the contribution $\Pi_{\gamma Z}(0)$ which vanishes for fermions
and sfermions but which is essential for the contribution of the
virtual $W$. In any case including gauge bosons in the
renormalisation of electromagnetic coupling requires the inclusion
of the $\Pi_{\gamma Z}(0)$ in Eq.~\ref{delgzuni} for gauge
invariance to be maintained\cite{grace-1loop}. We stress that we
will present the effect of the generalised effective coupling
$g_{\neuto \neuto Z}^{{\rm eff}}$ as an indication of the gauge
boson contribution while keeping in mind that this result may lead
to unitarity violation. Indeed through cutting rules, the $W$ loop
can be seen as made up of  the scattering $W^+W^- \to Z \to \mu^+
\mu^-$ that needs a compensation from the cut in the box shown in
Fig.~\ref{fig:loop_diags_neut1neut1mumu}(c). For the effective $Z
\mu^+ \mu^-$ coupling we only include the fermion/sfermion
contribution in Eqs.~\ref{delgzuni}-\ref{delsweff}, adding the
gauge bosons would require part of the 1-PI triangle contribution
to $Z \to \mu^+ \mu^-$.

\section{Analysis}

Since we will be studying different compositions of the
neutralinos we will take different values for the set
$M_1,M_2,\mu$. On the other hand the default parameters in the
Higgs sector are
\beqn
M_{A^0}=1{\rm TeV} \quad \quad \tb=4.
\eeqn
The sfermion sector is specified by a rather heavy spectrum (in
particular within the limits set by the LHC for
squarks\cite{lhc_susy_limits_june2011}). All sleptons left and
right of all generations have a common mass which we take  to be
different from the common mass in the squark sector. All
tri-linear parameters $A_f$ (including those for stops and
sbottom) are set to 0. The default values for the sfermion masses
are
\beqn
M_{{\tilde l}_{R}}&=&M_{{\tilde l}_{L}}=500{\rm GeV},\nonumber \\
M_{{\tilde u}_{R}}&=&M_{{\tilde d}_{R}}=M_{{\tilde Q}_{L}}=800{\rm
GeV},\nonumber \\
A_f&=&0 \,.
\eeqn
By default we  will focus on relatively light neutralinos (around
100 GeV) scattering with a relative velocity $v=0.2$.

To analyse consistently the efficiency of effective corrections we
will refer to the following quantities :

\beqn
\label{Delta_U} \Delta_{\rm eff}=\frac{\seff-\s0}{\s0} \,.
\eeqn
Here $\seff$ is the cross section calculated with the effective
couplings that include, by default, universal process independent particles
excluding gauge bosons and gauginos/higgsinos. We will explicitely specify when including all virtual particles in those corrections, referring to it as $\DU^{W}$. This correction
will  be compared to the correction solely due to the running of
the electromagnetic coupling, see Eq.~\ref{eq:running_alpha}. To
see how well the correction through the effective couplings
$\neuto f \tilde{f}$ and $\neuto \neuto Z$ reproduces the full
one-loop correction we introduce
\beqn
\Delta_{NU}=\frac{\sol-\seff}{\s0} \,, \nonumber \\
\Delta_{\rm full}=\frac{\sol-\s0}{\s0} \;.
\eeqn
with $\sol$  the full one-loop cross section, $\Delta_{NU}$
measures what we will refer to as the non-universal corrections
although strictly speaking this measures the remainder of all the
corrections that are not taken into account by the effective
vertices approach. $\Delta_{\rm full}=\Delta_{\rm
eff}+\Delta_{NU}$ is the full one-loop correction.

\subsection{Bino Case}
\subsubsection{Effective vs full corrections}
We first take $(M_1,M_2,\mu)=(90,200,-600)\ GeV$ which yields a
lightest  bino-like neutralino (the bino composition is 99\%) with
mass $\mneuto=91\ GeV$. At tree-level the cross section for
relative velocity $v=0.2$ is $\sigma_{\mu^+
\mu^-}^{\tilde{b}}=6.75 \times 10^{-3}{\rm pb}$. Note for further
reference that this is an order of magnitude larger than
annihilation into a pair of $W$'s: $\sigma_{W^+
W^-}^{\tilde{b}}=4.51 \times 10^{-4}{\rm pb}$. The annihilation
proceeds predominantly through the $t$-channel, binos coupling to
$Z$ are very much reduced. This leads to the following set of
corrections
\beqn
\DU=17.52\% (\DA=14.56\%) \quad \quad \DNU=2.06\%(\Delta_{\rm
full}=19.58\%).
\eeqn

\begin{figure}[htbp]
\begin{center}
\begin{tabular}{cc}
\includegraphics[width=0.5\textwidth]{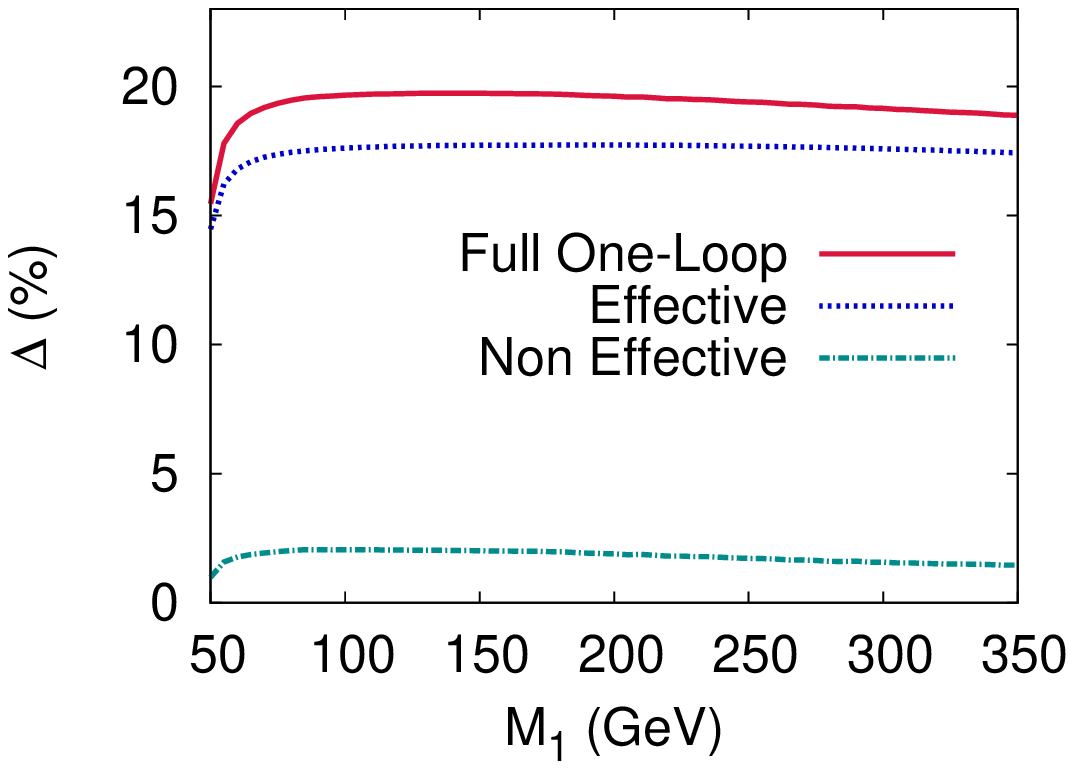}
&
\includegraphics[width=0.5\textwidth]{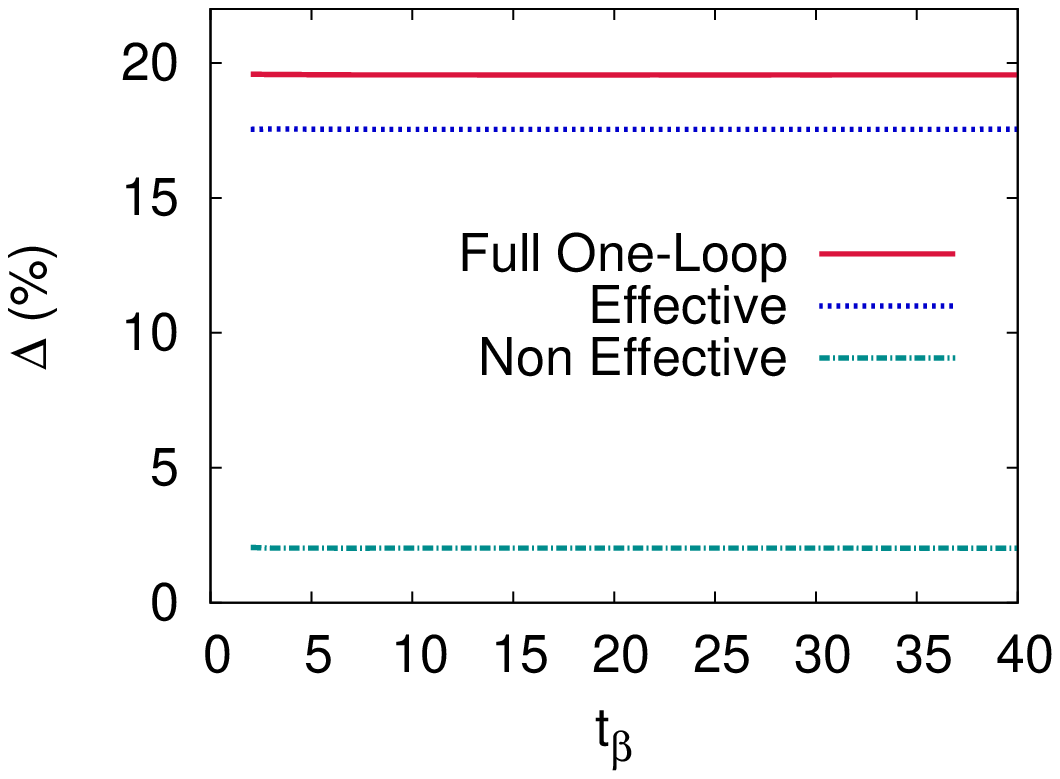}
\end{tabular}
\caption{{\em Corrections to the tree-level cross-section for the
process $\chii\chii\rightarrow \mu^+ \mu^-$ in the bino case as a
function of $M_1$ (left panel) and $\tb$ right panel. We show the
full one-loop, the effective correction and the difference which
we term non effective. $M_2=500, \mu=-600\ GeV$.}} \label{fig:bino2}
\end{center}
\end{figure}

For our first try  the effective universal coupling does
remarkably well falling short of only $2\%$ correction compared to
the full calculation. Note that although the most naive
implementation through a running of the electromagnetic coupling
fares also quite well it is nonetheless  $5\%$ off the total
correction, therefore the effective correction through the
effective couplings performs better. It must be admitted though
that
the bulk of the correction is through the running of $\alpha$. \\
To see how general this conclusion is we scanned over the set
$(M_1,M_2,\mu)$ while maintaining $\neuto$ with a 99\% bino like
component. This is simply obtained by taking $M_2=500,
\mu=-600\ GeV$ and scanning up to $M_1=350\ GeV$. We also checked how
sensitive our conclusion is depending on $\tb$ by varying $\tb$
from 2 to 40. The suspersymmtery breaking sfermion masses were
first left to their default values. As Fig.~\ref{fig:bino2} shows,
our conclusions remain quantitatively unchanged. There is no
appreciable dependence in $\tb$, we arrive at the same numbers as
our default $\tb$ value. As for the dependence in $M_1$ it is very
slight, for $M_1\sim 50\ GeV$ there is perfect matching with our
effective coupling implementation, then as $M_1$ increases to
$350\ GeV$, the non universal corrections remain negligible, below
$2\%$.

\begin{figure}[htbp]
\begin{center}
\includegraphics[width=0.8\textwidth]{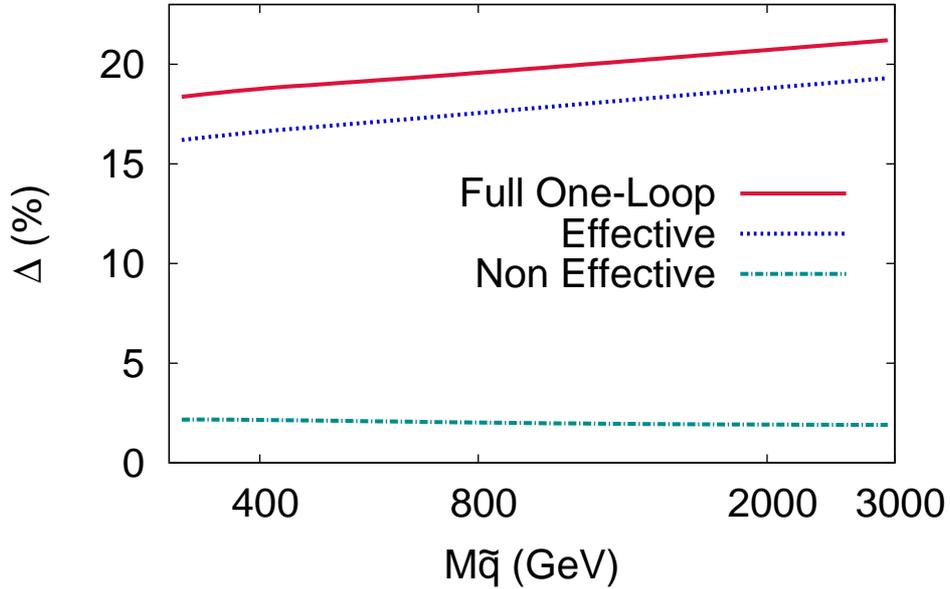}
\caption{{\em Corrections to the tree-level cross-section for the
process $\chii\chii\rightarrow\mu^+ \mu^-$ in the bino case
($M_1=90,M_2=200,\mu=-600\ GeV$) as function of the common soft
supersymmetry breaking squark mass.}} \label{fig:bino_nondecoup}
\end{center}
\end{figure}

The annihilation of neutralinos and hence the relic density  is a
very good example of the non decoupling effects of very heavy
sparticles, a remnant of supersymmetry breaking. The variation in
the fermion/sfermion masses is all contained in the effective
couplings that we have introduced. Leaving  the dependence on the
smuon mass at tree-level, and the very small (see below)
contribution of the smuon to the 1-PI vertex $\neuto \mu
\tilde{\mu}$, the bulk of the smuon mass dependence is within the
effective coupling. Fig.~\ref{fig:bino_nondecoup} shows how the
correction increases as the mass of the squarks increases from
$400\ GeV$ to $3\ TeV$, we take here a common mass for the
supersymmetry breaking squark masses (both right and left in all
three generations). The non universal correction of about $2\%$ is
insensitive to this change in squark masses whereas both $\DU$ and
$\DNU$ show the same logarithm growth that brings a $3\%$ change
as the squark mass is varied in the range $400\ GeV$ to $3\ TeV$.
This result also confirms that genuine vertex corrections and box
corrections are very small.\\
We have also extracted the individual
contribution of each species of fermions to the total
non-decoupling effect of sfermions. To achieve this we numerically
extracted the logarithm dependence of the non decoupling effect
for each species of sfermions. We have parameterised the effective
correction as
\beqn
\label{fit_delta_nd_eff} \Delta^f=a_{\tilde{f}} \ln
m_{\tilde{f}}/Q - a_f \ln m_f/Q + b_f \quad {\rm with} \quad
Q=2\mneuto \quad \tilde{f}=\tilde{d}_R+\tilde{u}_R+\tilde{Q}_L
\eeqn
The coefficients of the fit are given in
Table~\ref{table_af_bino}. As expected the fit to $a_f$ is
extremely well reproduced by the running of $\alpha$, {\it i.e},
$a_f=N_c q_f^2 \frac{4 \alpha}{3\pi}$. We also find
$a_{\tilde{e}}=a_{\tilde{\tau}}=a_{\tilde{\mu}},
b_e=b_\mu=b_\tau$. The fit to $a_f$ is made to validate the fit
procedure.
\begin{table}[htbp]
\begin{center}
$$
\begin{array}{|c|c|c|c|c|c|}
\cline{2-6}
\multicolumn{1}{c|}{} & a_{\tilde{Q}_L} & a_{\tilde{u}_R} & a_{\tilde{d}_R} & a_f & b_f\\
 \hline
e & 0.0010 & - & 0.00231 & 0.00310 & 0.15\% \\
(u,d)  & 0.000575 & 0.00236 & 0.000698 & (0.00413,0.00103) & 0.15\% \\
(t,b)  & -0.00406 & 0.00838 & 0.000661 & (0.00413,0.00103) & 0.16\%\\
\hline
\end{array}
$$
\end{center}
\caption{{\em Coefficients of the $\ln (m_f)$ (running couplings)
$\ln (m_{\tilde{f}})$(non decoupling effects) in $\DU$. $(c,s)$
give very similar results to $(u,d)$.}\label{table_af_bino}}
\end{table}
The most important observation is that the stops behave
differently, this is due to the Yukawa coupling of the top and
mixing. If there were not a compensation between left and right
contribution of the stops (compare to $\tilde{u}$) the
contribution of the stops would be even more important and would
dominate. Considering the different contributions and the scales
that enter our calculations it is difficult to attempt at giving an
analytical result, but leaving the stop aside the different
contributions to $a_{\tilde{f}}$ can be roughly approximated by
$y_f^2 N_c N_d/8/c_W^2$, $N_d=2$ for doublets and $1$ for singlet
of $SU(2)$. $y_f$ is the hypercharge, corresponding to the
couplings of the sfermions to the bino component.

\subsubsection{Scheme dependence in the bino case}
We have compared
the full correction to an approximate effective implementation and
observed that the approximation is quite good. However, even the
full correction, being computed at one-loop, it is potentially dependent on the renormalisation scheme chosen. As
discussed earlier we analyse the $\tb$ scheme dependence and the
$M_1$ scheme dependence. For $\tb$ we obtain the following
corrections:
$$ 19.58\% (DCPR), \quad 19.79\% (\o{DR}), \quad 19.51\% (MH).$$ This confirms that
the $\tb$ scheme dependence is very negligible. For the bino case
it is natural to reconstruct $M_1$ from the LSP, nonetheless
analysing the $M_1$ scheme dependence one chooses another
neutralino, say $\neutt$ which in our example is a wino-like. This
introduces more uncertainty or error since with this scheme the
corrections attain $24.08\%$, more than $4\%$ compared to the
usual scheme.

\subsection{Higgsino Case}

\subsubsection{Effective versus full corrections}
In the bino case our trial point had a neutralino of mass $91\ GeV$.
We therefore take the point (600,500,-100) which gives a LSP with
$M_{\chii}=95\ GeV$ with a  99\% higgsino content. The sfermion
parameters are the default values. In the higgsino case the cross
section is dominated by the exchange of the $Z$ in the
$s$-channel, so the bulk  of the corrections through the effective
couplings will be through the effective $\neuto \neuto Z$. For
further reference note that the tree-level cross section for
annihilation into muons is $\sigma_{\mu^+ \mu^-}^{\tilde{h}}=2.58
\times 10^{-3}{\rm pb}$, tiny and totally insignificant especially
compared to annihilation into $W$, $\sigma_{W^+
W^-}^{\tilde{h}}=18.83\ {\rm pb}$. This is an observation we will
keep in mind. The one-loop corrections we find for $\sigma_{\mu^+
\mu^-}^{\tilde{h}}$ are
\beqn
( {\rm for} \; \mu=-100{\rm GeV}) \quad \quad
\DU=13.55(\DA=14.62\%) \quad \DNU=-21.09\%(\Delta_{full}=-7.54\%)
\eeqn

This result is in a quite striking contrast to the bino case. The
effective coupling does not reproduce at all the full correction
and is off by as much as $21\%$. It looks like, at least for this
particular choice of parameters, that  going through the trouble
of implementing the effective $\neuto \neuto Z$ was in vain since
this correction is, within a per-cent, reproduced by the naive
running of $\alpha$. As we will see both these conclusions depend
much on the parameters of the higgsino and even the squark masses.
For example consider $\mu=-50\ GeV$, leaving all other parameters the
same. Of course this is a purely academic exercise, since in this
case, the charginos with mass $m_{\chi^\pm_1}=55\ GeV$ are ruled out
by LEP data. Nonetheless, in this case
\beqn
( {\rm for} \; \mu=-50{\rm GeV}) \quad \quad \DU=10.7(\DA=12\%)
\quad \DNU=-6.9\%(\Delta_{full}=3.8\%).
\eeqn
Had we included all particles in the effective vertex, we would get a correction $\DU^W=4.4\%$ improving thus the agreement with the one-loop correction for this particular
value of $\mu$ up to $0.6\%$. At the same
time a correction in terms of
a running of $\alpha$ will be off by more than 8\%. \\

\begin{figure}[htbp]
\begin{center}
\begin{tabular}{cc}
&
\includegraphics[width=0.8\textwidth]{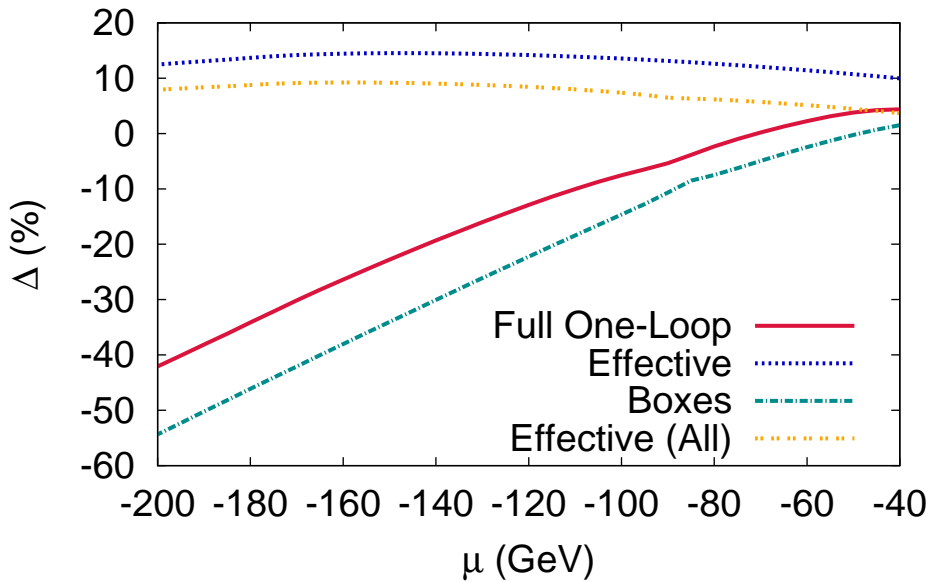}
\end{tabular}
\caption{{\em Corrections to the tree-level cross-section for the
process $\chii\chii\rightarrow \mu^+ \mu^-$ in the higgsino case
as a function of $\mu$. Shown are the effective vertex correction
(Effective, with only fermions/sfermions in the loops), the
effective $\neuto \neuto Z$ coupling  including all particles
denoted  Effective (All), the non QED boxes (Boxes) and the full
correction. $M_2=500, \mu=-600\ GeV$.}} \label{fig:hino2}
\end{center}
\end{figure}
\begin{figure}[htbp]
\begin{center}
\begin{tabular}{cc}
&
\includegraphics[width=0.7\textwidth]{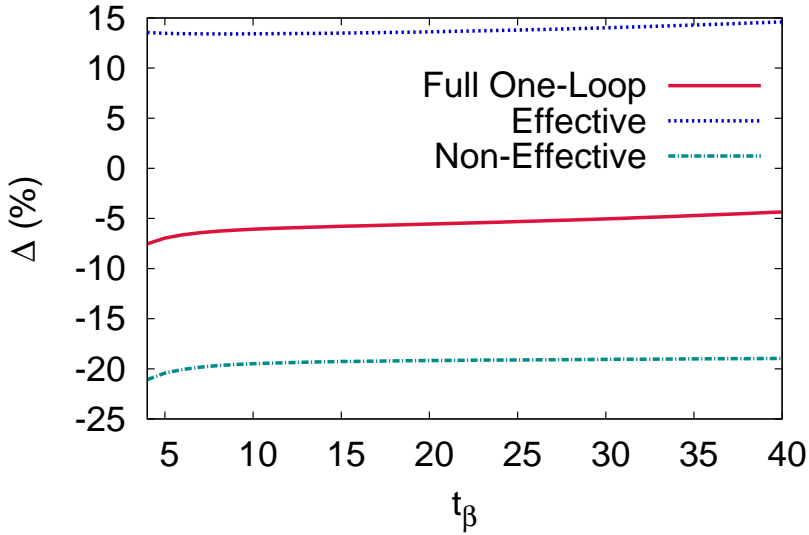}
\end{tabular}
\caption{{\em Corrections to the tree-level cross-section for the
process $\chii\chii\rightarrow \mu^+ \mu^-$ in the higgsino case
as a function of $\tb$. We show the full one-loop, the effective
correction and the remainder (Non-effective). $\mu=-100, M_2=500,
\mu=-600\ GeV$. }} \label{fig:hinotb}
\end{center}
\end{figure}

These two examples show that one can not, in the higgsino case,
draw a general conclusion on the efficiency of the effective
coupling as what was done in the bino case. Let us therefore look
at how the corrections change with $\mu$, and therefore with the
mass of the LSP,  while maintaining its higgsino nature. We have
varied $\mu$ from $-200\ GeV$ to $-40\ GeV$. Fig.~\ref{fig:hino2}
shows that the full correction is extremely sensitive to the value
of $\mu$. For $\mu=-200\ GeV$ the full one-loop correction is much
as $-42\%$, casting doubt on the loop expansion. The effective
coupling corrections with only fermions/sfermions on the other hand is much smoother and
positive bringing about $10\%$ correction. Including all particles
in the effective $\neuto \neuto Z$ vertex brings in an almost
constant  reduction of about $6\%$. Therefore as the value of
$|\mu|$ increases the effective one-loop corrections in the case
of the higgsino case can not  be trusted. The same figure shows
that the behaviour and the increase in the corrections is due
essentially to the contribution of the boxes. Here the boxes mean
the non QED box (involving an exchange of a photon which are
infrared divergent before including the real photon emission\footnote{The contribution of the QED box + real photon emission is only 0.1\%}). The
large contribution of the boxes can be understood by looking at
the box in Fig.~\ref{fig:tree_diags_neut1neut1mumu}(c). Indeed, as argued
previously, cutting through the box reveals that it represents
$\neuto \neuto \to W^+ W^-$ production that rescatter into $\mu^+ \mu^-$. Both these
process have a  very large cross sections compared to the
tree-level $\neuto \neuto \to \mu^+ \mu^-$. Our conclusion is
therefore that the effective vertex approximation is inadequate as
soon as the channel $\neuto \neuto \to W^+ W^-$ opens up. When
this occurs, in practical calculations of the relic density, the
channel $\neuto \neuto \to \mu^+ \mu^-$ is irrelevant and must
rather analyse the loop corrections to $\neuto \neuto \to W^+
W^-$. This process was studied in\cite{boudjema_chalons1,baro07} and will
be investigated further through an effective approximation in a
forthcoming study. \\
On the other hand, the dependence of the relative correction on
$\tb$ is quite modest even though there is certainly more
dependence than in the bino case, especially at lower values of
$\tgb$. This is shown in Fig.~\ref{fig:hinotb}.

\begin{figure}[htbp] \begin{center}
\mbox{\includegraphics[width=0.48\textwidth]{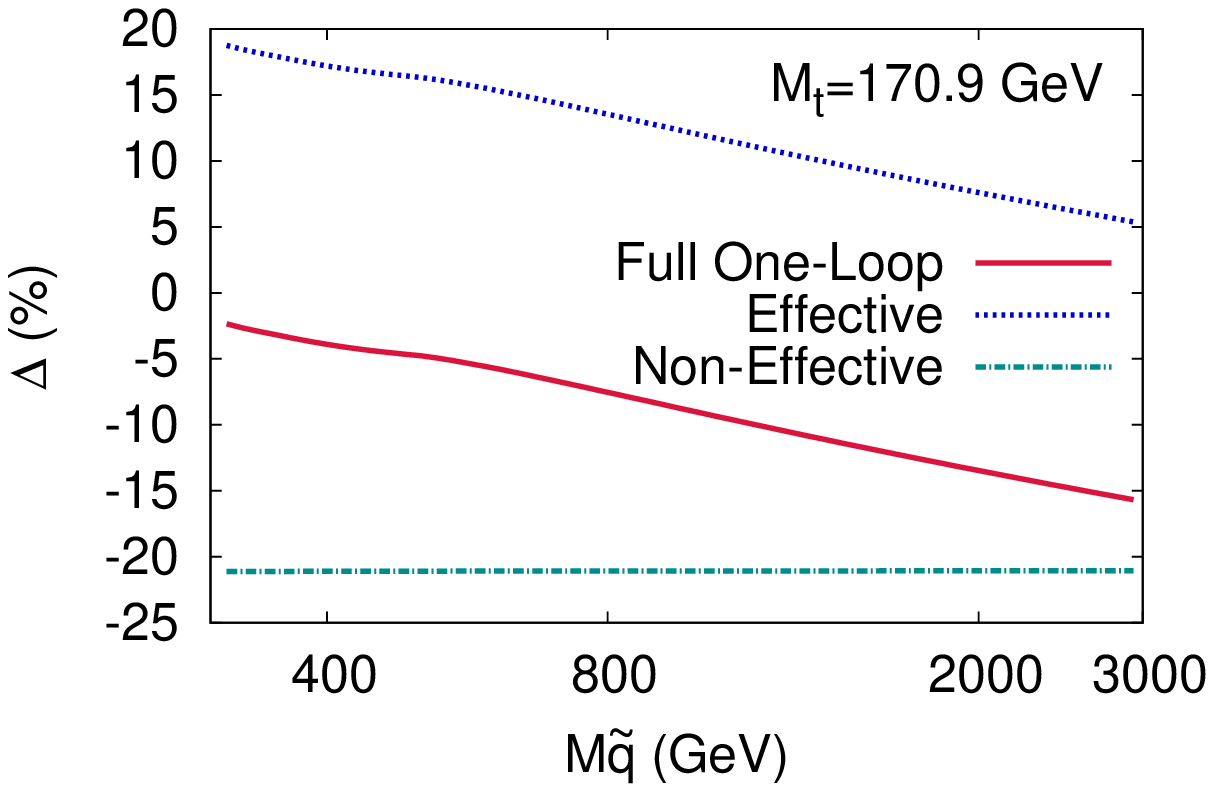}
\includegraphics[width=0.48\textwidth]{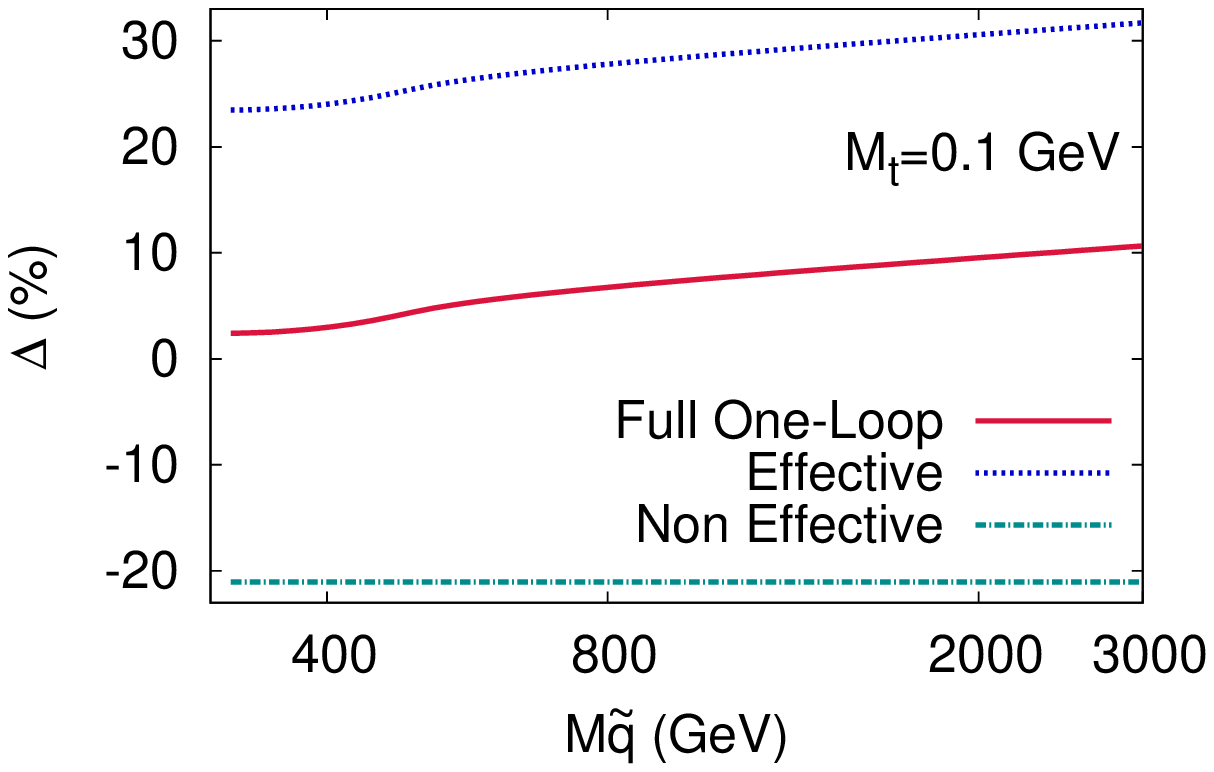}}
\caption{Corrections to the tree-level cross-section for the
process $\chii\chii \to \mu^+ \mu^-$ in the higgsino case
($M_1=600,M_2=500,\mu=-100$) as function of the common squark
mass. The right panel illustrates the case $m_t=0.1\ GeV$.}
\label{fig:higmsq}
\end{center}
\end{figure}

We now investigate the non-decoupling of very heavy squarks (and
heavy sfermions in general). Since we are in a Higgsino scenario
we expect the Yukawa of the fermions to play a more prominent role
than what was observed in the bino case. This is well supported by
our study. Fig.~\ref{fig:higmsq} shows how the effective (with
only fermions and sfermions) and the full correction gets modified
when the common mass of all squarks (all generations, left and
right) increases from $400\ GeV$ to $3\ TeV$. To
better illustrate the important effect of the Yukawa of the
top/stop sector we plot the corrections also for $m_t=0.1\ GeV$. For
$m_t=170.9\ GeV$, the correction drops by about $13\%$ when the mass
of the squarks increase from $400\ GeV$ to $3\ TeV$. This is much
more dramatic than in the bino case where we observed a 3\%
increase in the same range. Observe that for our default squark
mass of $800\ GeV$, the  effective correction including
sfermions/fermions is such that it almost accidently coincides
with the running of $\alpha$. If one switches off the top quark
mass, instead of a 13\% decrease we observe an 8\% increase for
$m_t=0.1\ GeV$! Observe that the difference one sees for
$m_{\tilde{Q}}=400\ GeV$ between $m_t=170.9\ GeV$ and $m_t=0.1\ GeV$ is
due essentially to the running  of $\alpha$ with very light top
that accounts for  $3\%$.

The special role played by the top can be seen even more clearly
from each individual contribution of the
fermion/sfermions and the fit of the contribution according to
Eq.~\ref{fit_delta_nd_eff} as was done for the bino case.
\begin{table}[htbp]
\begin{center}
$$
\begin{array}{|c|c|c|c|c|c|}
\cline{2-6}
\multicolumn{1}{c|}{}& a_{\tilde{Q}_L} & a_{\tilde{u}_R} & a_{\tilde{d}_R} & a_f & b_f\\
 \hline
e  & 0.00304 & - & 0.000366 & 0.00309 & -0.12\% \\
(u,d)  & 0.00861 & 0.000489 & 0.000122 & (0.00414,0.00101) & -0.13\% \\
(t,b) & -0.0701 & 0.000826 & 0.000108 & (0.00414,0.00101) & 0.13\%\\
\hline
\end{array}
$$
\end{center}
\caption{{\em Coefficients of the $\ln (m_f)$ (running couplings)
$\ln (m_{\tilde{f}})$(non decoupling effects) in $\DU$. $(c,s)$
give very similar results to $(u,d)$. Higgsino
case.}\label{table_af_higgsino}}
\end{table}
The contribution of the stop is clearly (especially through
$\tilde{Q}_L$) an order of magnitude larger than for all other
sfermions, see Table~\ref{table_af_higgsino}. It is the only one
that brings a negative contribution. Since this effect is in the
universal $\neuto \neuto Z$ it will show up in many processes
where the higgsino contributes.

\subsubsection{Scheme dependence in the higgsino case}
We analyse here the $\tb$ scheme dependence and the $M_1$ scheme
dependence. For $\tb$ we obtain the following corrections:
$$ -7.5\% (DCPR), \quad -12.4\% (\o{DR}), \quad -4.76\% (MH).$$
As expected and in line with the behaviour of the corrections with
respect to $\tgb$, Fig.~\ref{fig:hinotb}, we see that the
corrections though larger than in the bino case are nonetheless
within 5\%. On the other hand, expectedly the choice of $M_1$ has
less impact than in the bino case where the reconstruction of
$M_1$ is essential to define the LSP. In the case of the higgsino,
changing the $M_1$ scheme turns the full correction from -7.5\% (in DCPR
scheme for $\tb$) to -10.7\%, a 3\% uncertainty.

\section{Conclusions}

Very few analyses have been done taking into account the full
one-loop corrections to the annihilation cross sections entering
the computation of the relic density despite the fact that this
observable is now measured within 3\% precision. In supersymmetry
radiative corrections have been known to be important, yet
practically all analyses that constrain the parameter space of
supersymmetry are performed with tree-level annihilation cross
sections. Taking into account the full one-loop corrections to a
plethora of processes is most probably unrealistic. On the other
hand one must incorporate, if possible simply and quickly, a
parameterisation of the theory error or implement the corrections
through effective couplings of the neutralino, in the case of
supersymmetry. This is what we have attempted in this study for
two of the most important couplings of the neutralinos $\neuto f
\tilde{f}$ and $\neuto \neuto Z$. In order to look more precisely
at the impact of each of these effective couplings we take as a
testing ground a most simple process, $\neuto \neuto \to \mu^+
\mu^-$ and select a neutralino that is either almost pure bino or
pure higgsino. We do not strive at finding a scenario with the
correct relic density since our primary task is to study this
vertices and the approximations in detail. In this exploratory
study taking a final state involving gauge bosons would only
confuse the issues. Nonetheless, the impact of the gauge bosons is
studied. Indeed, we have shown how the construction of the
effective $\neuto \neuto Z$ is quite different from that of the
$\neuto f \tilde{f}$. For the latter the effective coupling
involves self-energy corrections, whereas for the former the
one-particle irreducible vertex correction must be added. These
examples and the construction of the effective coupling already
pave the way to a generalisation to the effective couplings
$\neuto \neuto h,H,A$ and $\neuto \chi^+ W$ which we will address
in forthcoming publications with applications to different
process, including gauge boson final states. Even with the
effective couplings we have derived, we could generalise the study
of $\neuto \neuto \to \mu^+ \mu^-$ to cover not only pure winos,
but also mixed scenarios and also heavy fermions. \\
Our preliminary study on the simple process $\neuto \neuto \to
\mu^+ \mu^-$ is already very instructive. To summarise the bino
case, we can state that the effective couplings approach is a very
good approximation that embodies extremely well the non decoupling
effects from heavy sfermions, irrespective of many of the
parameters that are involved in the calculation, as long as one is
in an almost pure bino case. The effective coupling implementation is
within $2\%$ of the full one-loop calculation. Here, this reflects
essentially the correction to the $\neuto f \tilde{f}$ coupling.
The scheme dependence from $\tb$ is very small, this result
stands for large $M_1$ masses as long as the neutralino is more
than $90\%$ bino like. In particular for higgsino-like
LSP in excess of $90\ GeV$ as imposed by present limits on the
chargino, the effective coupling implementation in the
annihilation $\neuto \neuto \to \mu^+ \mu^-$ fails. It worsens as
the mass increases due to the importance of a large box
contribution corresponding to the opening up of $\neuto \neuto \to
W^+ W^-$ which would in any case be  the dominant process to take
into account when calculating the relic density. The large Yukawa
of the top has a big impact on the radiative corrections and in
particular on the non-decoupling contribution of a very heavy
stop. Although this is an example which shows, in principle, the
failure of the effective approach apart from correctly reproducing
 the non-decoupling effect of very heavy squarks, we need
 further investigation on the dominant processes, in this case
 annihilations into $W,Z$, to see if these dominant processes
 could on the other hand be reproduced by an effective coupling
 approach. If the effective approach turns out to be efficient
 for the dominant processes, where and if the box corrections are tamed, the effective coupling could still be a good alternative for the
 calculation of the relic density with high precision. We leave
 many of these interesting issues to further analyses.

\vspace*{3cm}
{\bf \large Acknowledgments}\\
We would like to thank Guillaume Chalons for many useful discussions.  This work is part of the
French ANR project, ToolsDMColl BLAN07-2 194882 and is supported in part by the GDRI-ACPP
of the CNRS (France). GDLR held a fellowship from la Région Rhones-Alpes. This work was supported by TRR33 "The Dark Universe".


\begin{thebibliography}{10}

\bibitem{lhc_susy_limits_june2011}
see for example, S.~Caron for the ATLAS collaboration, arXiv:1106.1009
  [hep-ex].\\ J.~B.~G.~da Costa {\it et al.} [Atlas Collaboration], Phys.\
  Lett.\ B {\bf 701} (2011) 186 [arXiv:1102.5290 [hep-ex]]. \\ G.~Aad {\it et
  al.} [ATLAS Collaboration], Eur.\ Phys.\ J.\ C {\bf 71} (2011) 1682
  [arXiv:1103.6214 [hep-ex]].\\ S.~Chatrchyan {\it et al.} [ CMS Collaboration
  ],[arXiv:1107.1279 [hep-ex]].

\bibitem{lhc_higgs_limits_june2011}
See for example, W.~Murray, {\em Higgs searches at the LHC}, Plenary Summary
  talk at the Europhysics Conference on High-Energy Physics, Grenoble (2011).

\bibitem{Jarosik:2010iu}
N.~Jarosik {\it et al.}, Astrophys.\ J.\ Suppl.\ {\bf 192} (2011) 14
  [arXiv:1001.4744 [astro-ph.CO]].

\bibitem{Percival:2009xn}
B.~A.~Reid {\it et al.} [SDSS Collaboration], Mon.\ Not.\ Roy.\ Astron.\ Soc.\
  {\bf 401} (2010) 2148 [arXiv:0907.1660 [astro-ph.CO]].

\bibitem{Riess:2009pu}
A.~G.~Riess {\it et al.}, Astrophys.\ J.\ {\bf 699} (2009) 539 [arXiv:0905.0695
  [astro-ph.CO]].

\bibitem{Komatsu:2010fb}
E.~Komatsu {\it et al.} [WMAP Collaboration], Astrophys.\ J.\ Suppl.\ {\bf 192}
  (2011) 18 [arXiv:1001.4538 [astro-ph.CO]].

\bibitem{micromegas}
G.~B\'{e}langer, F.~Boudjema, A.~Pukhov, A.~Semenov, \textit{Comput. Phys.
  Commun.} {\bf 149} (2002) 103, hep-ph/0112278;\\ G.~B\'elanger, F.~Boudjema,
  A.~Pukhov, A.~Semenov, \textit{Comput. Phys. Commun.} {\bf 176} (2007) 367,
  hep-ph/0607059;\\ G.~B\'elanger, F.~Boudjema, A.~Pukhov, A.~Semenov,
  \textit{Comput. Phys. Commun.} {\bf 174} (2006) 577, hep-ph/0405253;\\ {\tt
  http://lapth.in2p3.fr/micromegas}.

\bibitem{darksusy}
{\tt DarkSUSY}: P.~Gondolo \textit{et al.}, \textit{JCAP} {\bf 0407} (2004)
  008, astro-ph/0406204;\\ {\tt http://www.physto.se/$\sim$edsjo/darksusy/}.

\bibitem{superiso-relic}
{\tt SuperIso Relic}: A.~Arbey, F.~Mahmoudi, A.~Arbey and F.~Mahmoudi, Comput.\
  Phys.\ Commun.\ {\bf 181} (2010) 1277 [arXiv:0906.0369 [hep-ph]]. \\ Comput.\
  Phys.\ Commun.\ {\bf 182}, 1582 (2011).\\ {\tt
  http://superiso.in2p3.fr/relic/}.

\bibitem{baro07}
N.~Baro, F.~Boudjema, A.~Semenov, \textit{Phys. Lett.} {\bf B660} (2008) 550,
  arXiv:0710.1821 [hep-ph].

\bibitem{Freitas-relic-qcd}
A.~Freitas, Phys.\ Lett.\ B {\bf 652} (2007) 280 [arXiv:0705.4027 [hep-ph]].

\bibitem{Klasen-relic-qcd}
B.~Herrmann, M.~Klasen, \textit{Phys. Rev.} \textbf{D76} (2007) 117704,
  arXiv:0709.2232 [hep-ph]. \\ B.~Herrmann, M.~Klasen, K.~Kovarik,
  \textit{Phys. Rev.} \textbf{D79} (2009) 061701, arXiv:0901.0481 [hep-ph]. \\
  B.~Herrmann, M.~Klasen, K.~Kovarik, \textit{Phys. Rev.} \textbf{D80} (2009)
  085025, arXiv:0907.0030[hep-ph].

\bibitem{boudjema_chalons1}
N.~Baro, F.~Boudjema, G.~Chalons and S.~Hao, Phys.\ Rev.\ D {\bf 81} (2010)
  015005 [arXiv:0910.3293 [hep-ph]]. 

\bibitem{Bjorn_review_rc}
For a recent review, see B.~Herrmann, arXiv:1011.6550 [hep-ph].

\bibitem{boudjema_gondolo}
F.~Boudjema, J.~Edsjo and P.~Gondolo, in {\em Particle dark matter} 325-344,
  Oxford University Press (2010) G.~Bertone, editor; 
  Matter and at the Colliders,'' [arXiv:1003.4748 [hep-ph]].

\bibitem{feng_nondecoupling}
H.~C.~Cheng, J.~L.~Feng and N.~Polonsky, Phys.\ Rev.\ D {\bf 56} (1997) 6875,
  [arXiv:hep-ph/9706438]; idem Phys.\ Rev.\ D {\bf 57} (1998) 152,
  [arXiv:hep-ph/9706476]. 

\bibitem{randall_nondecoupling}
E.~Katz, L.~Randall and S.~f.~Su, Nucl.\ Phys.\ B {\bf 536} (1998) 3
  [arXiv:hep-ph/9801416].

\bibitem{nojiri_nondecoupling}
S.~Kiyoura, M.~M.~Nojiri, D.~M.~Pierce and Y.~Yamada, Phys.\ Rev.\ D {\bf 58},
  075002 (1998) [arXiv:hep-ph/9803210].

\bibitem{Sommerfeldinclude}
For a recent review see, A.~Hryczuk, Phys.\ Lett.\ B {\bf 699} (2011) 271
  [arXiv:1102.4295 [hep-ph]].

\bibitem{delta_mb}
L.~J.~Hall, R.~Rattazzi and U.~Sarid, Phys.\ Rev.\ D {\bf 50} (1994) 7048
  [arXiv:hep-ph/9306309]. \\ M.~S.~Carena, M.~Olechowski, S.~Pokorski and
  C.~E.~M.~Wagner, Nucl.\ Phys.\ B {\bf 426} (1994) 269 [arXiv:hep-ph/9402253].
  \\ M.~S.~Carena, D.~Garcia, U.~Nierste and C.~E.~M.~Wagner, Nucl.\ Phys.\ B
  {\bf 577} (2000) 88 [arXiv:hep-ph/9912516].

\bibitem{nonconventional-relic}
P.~Salati, Phys.\ Lett.\ B {\bf 571} (2003) 121 [arXiv:astro-ph/0207396].\\
  S.~Profumo and P.~Ullio, JCAP {\bf 0311} (2003) 006 [arXiv:hep-ph/0309220].\\
  F.~Rosati, Phys.\ Lett.\ B {\bf 570} (2003) 5 [arXiv:hep-ph/0302159].\\
  C.~Pallis, JCAP {\bf 0510} (2005) 015 [arXiv:hep-ph/0503080].\\ G.~B.~Gelmini
  and P.~Gondolo, Phys.\ Rev.\ {\bf D74} (2006) 023510
  [arXiv:hep-ph/0602230].\\ D.~J.~H.~Chung, L.~L.~Everett, K.~Kong and
  K.~T.~Matchev, arXiv:0706.2375 [hep-ph].\\ M.~Drees, H.~Iminniyaz and
  M.~Kakizaki, \textit{Phys.Rev.} {\bf D76} (2007) 103524,
  [arXiv:0704.1590[hep-ph]].\\ A.~Arbey and F.~Mahmoudi, JHEP {\bf 1005} (2010)
  051 [arXiv:0906.0368 [hep-ph]].

\bibitem{lanhep}
A. Semenov. {\it \texttt{LanHEP} --- a package for automatic generation of Feynman
  rules. User's manual.}; hep-ph/9608488. \\ A.~Semenov, \textit{Nucl. Inst.
  Meth. and Inst.} {\bf A393} (1997) 293;\\ A.~Semenov, \textit{Comp. Phys.
  Commun.} {\bf 115} (1998) 124;\\ A.~Semenov, hep-ph/0208011; \\ A.~Semenov,
  {\em Comput. Phys. Commun.} {\bf 180} (2009) 431, arXiv:0805.0555 [hep-ph].

\bibitem{feynarts}
J.~K\"ublbeck, M.~B\"ohm, A.~Denner, \textit{Comp. Phys. Commun.} {\bf 60}
  (1990) 165;\\ H.~Eck, J.~K\"ublbeck, \textit{Guide to FeynArts~1.0},
  W\"urzburg, 1991;\\ H.~Eck, \textit{Guide to FeynArts~2.0}, W\"urzburg,
  1995;\\ T.~Hahn, \textit{Comp. Phys. Commun.} {\bf 140} (2001) 418,
  hep-ph/0012260.

\bibitem{formcalc}
T.~Hahn, M.~Perez-Victoria, \textit{Comp. Phys. Commun.} {\bf 118} (1999) 153,
  hep-ph/9807565;\\ T.~Hahn, hep-ph/0406288; hep-ph/0506201.

\bibitem{looptools}
T. Hahn, {\tt LoopTools}, \verb+http://www.feynarts.de/looptools/+.

\bibitem{boudjema_temes}
F.~Boudjema, A.~Semenov, D.~Temes, \textit{Phys. Rev.} {\bf D72} (2005) 055024,
  hep-ph/0507127.

\bibitem{Sloops-higgspaper}
N.~Baro, F.~Boudjema, A.~Semenov, \textit{Phys. Rev.} {\bf D78} (2008) 115003,
  arXiv:0807.4668 [hep-ph].

\bibitem{baro09}
N.~Baro, F.~Boudjema, \textit{Phys. Rev} {\bf D80} (2009) 076010,
  [arXiv:0906.1665 [hep-ph]].

\bibitem{grace-1loop}
G.~B\'{e}langer, F.~Boudjema, J.~Fujimoto, T.~Ishikawa, T.~Kaneko, K.~Kato,
  Y.~Shimizu, \textit{Phys. Rep.} {\bf 430} (2006) 117, hep-ph/0308080.

\bibitem{eennhletter}
G. B\'{e}langer, F. Boudjema, J. Fujimoto, T. Ishikawa, T. Kaneko, K. Kato and
  Y. Shimizu, {\em Phys. Lett. } {\bf B559} (2003) 252; hep-ph/0212261.

\bibitem{DabelsteinHiggs}
A.~Dabelstein, \textit{Z. Phys.} {\bf C67} (1995) 495, hep-ph/9409375.

\bibitem{DCPR}
P.H.~Chankowski, S.~Pokorski and J.~Rosiek, \textit{Nucl. Phys.} {\bf B423}
  (1994) 437, hep-ph/9303309.

\bibitem{CDR}
D.Z.~Freedman, K.~Johnson, J.I.~Latorre, \textit{Nucl. Phys.} {\bf B371} (1992)
  353; \\ P.E.~Haagensen, \textit{Mod. Phys. Lett.} {\bf A7} (1992) 893,
  hep-th/9111015;\\ F.~del~Aguila, A.~Culatti, R.~Mu\~noz Tapia,
  M.~P\'erez-Victoria, \textit{Phys. Lett.} {\bf B419} (1998) 263,
  hep-th/9709067;\\ F.~del~Aguila, A.~Culatti, R.~Mu\~noz Tapia,
  M.~P\'erez-Victoria, \textit{Nucl. Phys.} {\bf B537} (1999) 561,
  hep-ph/9806451;\\ F.~del~Aguila, A.~Culatti, R.~Mu\~noz Tapia,
  M.~P\'erez-Victoria, \textit{Nucl. Phys.} {\bf B504} (1997) 532,
  hep-ph/9702342.

\bibitem{Hollik_susyeff}
J.~Guasch, W.~Hollik and J.~Sola, JHEP {\bf 0210}, 040 (2002)
  [arXiv:hep-ph/0207364]. 

\end{thebibliography}
\end{document}